\documentclass[12pt,preprint]{aastex}
\begin{document}

\baselineskip=0.8cm

\parindent=1.0cm

\title{Seeing Red in M32: Constraints on the Stellar Content from Near and 
Mid-Infrared Observations and Applications for Studies 
of More Distant Galaxies \altaffilmark{1}\altaffilmark{2}\altaffilmark{3}}

\author{T. J. Davidge}

\affil{Dominion Astrophysical Observatory,
\\National Research Council of Canada, 5071 West Saanich Road,
\\Victoria, BC Canada V9E 2E7}

\altaffiltext{1}{Based on observations obtained at the Gemini Observatory, which is
operated by the Association of Universities for Research in Astronomy, Inc., under a
cooperative agreement with the NSF on behalf of the Gemini partnership: the National
Science Foundation (United States), the National Research Council (Canada), CONICYT
(Chile), the Australian Research Council (Australia), Minist\'{e}rio da Ci\^{e}ncia,
Tecnologia e Inova\c{c}\~{a}o (Brazil) and Ministerio de Ciencia, Tecnolog\'{i}a e
Innovaci\'{o}n Productiva (Argentina).}

\altaffiltext{2}{This research used the facilities of the Canadian Astronomy Data 
Centre operated by the National Research Council of Canada with the support 
of the Canadian Space Agency.}

\altaffiltext{3}{This research has made use of the NASA/IPAC Infrared Science Archive, 
which is operated by the Jet Propulsion Laboratory, California Institute of Technology, 
under contract with the National Aeronautics and Space Administration.}

\begin{abstract}

	The properties of asymptotic giant branch (AGB) stars in the Local Group galaxy 
M32 are investigated using ground and space-based observations that span the 
$1 - 8\mu$m wavelength interval, with the goal of demonstrating the utility of 
infrared observations as probes of stellar content. Comparisons with 
isochrones indicate that the brightest resolved stars in M32 have ages of a few Gyr, 
and are younger on average than AGB stars with the same intrinsic brightness in 
the outer disk of M31. Accounting for stellar variability is shown to be essential 
for modelling AGB luminosity functions (LFs). Model LFs that assume the star-forming 
history measured by Monachesi et al. (2012, ApJ, 745, 97) and the variability 
properties of Galactic AGB stars match both the $K$ and [5.8] LFs of M32. 
Models also suggest that the slope of the [5.8] LF between M$_{[5.8]} = -8.5$ and 
--10.0 is sensitive to the mix of stellar ages, and a sizeable fraction 
of the stars in M32 must have an age older than 7 Gyr in order to match 
the [5.8] LF. The structural properties of M32 in the infrared are also investigated. 
The effective radii that are computed from near-infrared and mid-infrared isophotes
are similar to those measured at visible wavelengths, suggesting that the 
stellar content of M32 is well-mixed. However, isophotes at radii $> 16$ arcsec 
($> 60$ parcsecs) in the near and mid-infrared are flatter 
than those at visible wavelengths. The coefficient of the fourth-order cosine term in 
the fourier expansion of isophotes changes from `boxy' values at 
$r < 16$ arcsec to `disky' values at $r > 48$ arcsec in [3.6] and [4.5]. 
The mid-infrared colors near the center of M32 do not vary 
systematically with radius, providing evidence of a well-mixed 
stellar content in this part of the galaxy.

\end{abstract}


\section{INTRODUCTION}

\subsection{Asymptotic Giant Branch Stars and M32}

	The observational properties of stars near 
the main sequence turn-off (MSTO) have a natural primacy in studies 
of the star-forming history (SFH) of galaxies. However, 
with the exception of stars that formed within the past few hundred Myr, 
crowding and the intrinsic faintness of the MSTO restrict its use 
to Local Group galaxies. Core helium-burning stars provide an alternate means of 
probing the stellar content of more distant systems (e.g. McQuinn et al. 2011), 
although these stars can only be resolved in relatively low 
surface brightness areas of galaxies in the nearest groups. 

	Stars evolving near the tip of the Asymptotic Giant Branch (AGB) are among the 
most luminous objects in a galaxy, and so offer another means of 
investigating stellar content. The stars on the 
upper few magnitudes of the AGB can have a wide 
range of ages, extending back $10+$ Gyr for solar 
metallicities, and so have the potential to provide constraints on the 
SFH over a large fraction of the age of the Universe. The red colors of the most 
luminous AGB stars make them well-suited for study with adaptive 
optics (AO)-equipped ground-based telescopes, since the 
corrections delivered by AO systems improve towards longer wavelengths. 

	Because they are very bright and have colors that are much 
redder than the majority of other luminous stars in a galaxy, 
AGB stars will be resolved in the near and mid-infrared with future space-based 
facilites -- such as the JWST -- in systems that are well outside of the Local Group.
Still, given their highly evolved nature, models of AGB stars
can be uncertain (e.g. discussion by Marigo et al. 2008), and there has been mixed 
success matching observations with model predictions (e.g. Zibetti et al. 2013; 
Melbourne et al.  2012; Girardi et al. 2010). This has motivated recent efforts to 
characterize the behaviour of AGB stars in dwarf galaxies that span a broader range of 
metallicities than is sampled by the Magellanic Clouds and the Milky-Way, which 
have heretofore been the key calibrators of AGB properties (e.g. Rosenfield 
et al. 2014; Melbourne et al. 2010).

	The Local Group compact elliptical (cE) galaxy 
M32 is a fundamental laboratory for probing AGB evolution 
and investigating the role that these stars play 
in stellar content studies. M32 is close enough to allow 
the MSTO stars that probe the early phases of its evolution to be resolved, 
so that its SFH can be probed in detail (e.g. Monachesi et al. 2012). At the same time, 
M32 is also far enough away and has a high enough stellar density that integrated 
spectra that are not subject to stochastic sampling effects can be recorded 
in its high surface brightness central regions. That a rich population of 
AGB stars in M32 can be resolved over a wide range of radii (e.g. 
Davidge \& Jensen 2007) also means that the problems associated with 
low number statistics that plague studies of AGB stars in more diffuse and/or 
less massive systems are avoided. 

	The utility of M32 for testing techniques to probe stellar content has 
been amply demonstrated in the past. M32 has been the target of numerous 
spectroscopic studies, and these have revealed signatures of a substantial -- 
possibly even dominant -- population with an age of a few Gyr. 
Absorption lines originating from the Balmer series of Hydrogen 
-- most noteably H$\beta$ (e.g. O'Connell 1980; Bica et al. 1990; Rose 
1994; Schiavon et al. 2004) -- and the first-overtone CO bands (Davidge 1990) 
have depths in the integrated spectrum of M32 that are indicative of a 
large intermediate age population. This has been confirmed by the subsequent 
study of resolved AGB (e.g. Davidge 2000; Davidge \& Jensen 2007) and MSTO (Monachesi 
et al. 2012) stars. Spectroscopic studies also reveal a near solar 
luminosity-weighted metallicity. This is consistent with the metallicities of 
individual AGB stars in the central regions of M32 (Davidge et al. 2010) and the 
mean photometric properties of red giant branch (RGB) stars at large 
radii (Grillmair et al. 1996; Monachesi et al. 2012). 

\subsection{The Evolution of M32}

	The origins of M32 are enigmatic. 
While the structural characterics of M32 are reminiscent of a 
classical elliptical galaxy (e.g. Kormendy 1985), its near-solar 
metallicity is consistent with it being the remnant of what was once a much 
larger galaxy. In addition to having a relatively high mean 
metallicity for its mass, the location of M32 on the Fe $vs.$ Mg$_2$ plane 
is indicative of a [Mg/Fe] that is lower than in most early-type galaxies (Worthey 
et al. 1992). This hints that SNe I played a significant role in 
the chemical enrichment of M32. This is not expected for 
spheroidal systems that experienced rapid chemical enrichment, but is associated 
with disks that are stable over time scales of at least a few hundred Myr.

	M32 and M31 are in close physical proximity at present, with a projected 
separation of only a few kpc (e.g. Evans et al. 2000; Sarajedini et al. 
2012). This raises the prospect of past interactions between them, and 
Bekki et al. (2001) suggest that M32 is the remnant of 
a disk-dominated galaxy that was threshed by tidal interactions with M31. 
If the progenitor of M32 was a disk galaxy then this could explain the chemical 
mixture of M32. The tidal threshing model could also explain 
other properties of M32, such as the lack of a globular cluster system, its low gas 
content (Sage et al. 1998; Grcevich \& Putman 2009), and the uniform mixing of RR 
Lyraes with other stellar types (Sarajedini et al. 2012).

	There may have been more recent encounters between M31 and M32. Block et al. 
(2006) argue that a satellite of M31 -- presumably M32 -- passed through the inner 
regions of its disk a few hundred Myr ago. Such an encounter is consistent with 
the complex dynamical state of the ISM near the center of M31 (e.g. Melchior \& Combes 
2011). Smith et al. (2012) find that the dust properties of the central few kpc of 
M31 may also have been affected by an interaction, and note that the 
dust mass within the central 3 kpc is comparable to that expected in a dwarf galaxy.
The concentration of bright AGB stars along one arm of the M31 major axis 
(Davidge et al. 2012; Davidge 2012; Dalcanton et al. 2012) may also have formed during 
this event. A basic prediction of the Block et al. (2006) model 
is that the interaction with M32 triggered disturbances in the ISM 
of M31 that are propogating outwards, spurring star-formation as they go. The presence 
of such a propogating density wave is not consistent with the long-lived nature 
of the star-forming rings (Davidge et al. 2012; Dalcanton et al. 2012), and 
may have difficulties explaining related shell-like structures in the outermost 
regions of the disk (Fritz et al. 2012). 

	Monachesi et al. (2012) use deep images of the outer regions of M32 
to resolve stars near the MSTO and examine the SFH. They conclude that 
M32 is not coeval. There is a dominant component that is older than 5 Gyr, and 
a significant population with an age 2 -- 5 Gyr. The MSTO of the oldest stars 
is not detected, and so firm constraints can not be placed 
on the mass fraction that formed during very early epochs. A population of stars that 
is younger than 2 Gyr and accounts for a few percent of the total stellar mass 
was also detected. While MSTO stars are primary probes of stellar age, blue 
stragglers, and contamination from clusters of young stars in the outer disk of M31 add 
uncertainties to age estimates obtained from the MSTO, and may account for 
at least some fraction of this relatively young component.

\subsection{The Present Study}

	Obtaining agreement between the SFHs deduced from AGB and 
MSTO stars in nearby galaxies is an important step if 
AGB stars are to be considered as probes of more distant systems. 
There are challenges inherent to detecting the most evolved 
AGB stars in the visible, even though this is a wavelength region 
that has been explored in detail and is where space-based 
facilities typically deliver their best angular resolution. The most 
evolved RGB and AGB stars with moderate and higher metallicities have cool photospheric 
temperatures that are conducive to the formation of 
deep absorption bands of TiO and ZrO at visible and red wavelengths, 
and blanketing from these species forces the RGB and 
AGB sequences in the color-magnitude diagrams (CMDs) at visual wavelengths to slump 
over (e.g. see discussion in Ortolani et al. 1990). This complicates efforts to measure 
the luminosities of the brightest AGB stars, while also making AGB-tip stars more 
susceptible to blending with sources that have lower intrinsic luminosities. The 
detection of highly evolved AGB stars at visible wavelengths may also be affected by 
obscuration from circumstellar dust. 

	Many of the challenges that are inherent to the detection of bright AGB stars 
at visible wavelengths are eased when observing in the near- and mid-infrared 
(NIR and MIR). The effects of line blanketing and circumstellar 
obscuration are significantly reduced at these wavelengths. 
In addition, cool AGB stars are significantly brighter 
in the infrared than the bulk of the underlying population that contributes most of the 
stellar mass. When observed in the NIR and MIR, the coolest AGB stars are thus 
less susceptible to blending and can be used to probe regions in galaxies where MSTO 
stars can not be detected. In the particular case of M32, AGB stars have been resolved 
in $H$ and $K$ by AO-equipped ground-based telescopes to within a few arcsec 
of the galaxy center (e.g. Davidge et al. 2000, Davidge \& Jensen 2007; 
Davidge et al. 2010). This is a part of the galaxy where there is no hope of resolving 
stars near the MSTO in the foreseeable future.

	The high emissivity associated with ground-based facilities is a formidable 
obstacle for studying all but the brightest sources in the MIR. While deep MIR 
observations are the exclusive purview of space-based facilities, these 
same facilities have poorer angular resolution than ground-based facilities due 
to their modest apertures. The superior angular resolution offered by 
ground-based facilities provides a niche that can be exploited, at least at the 
shortest MIR wavelengths.

	In the present study, data obtained with the 
Canada-France-Hawaii Telescope (CFHT), Gemini North (GN), and the SPITZER satellite 
are used to investigate the NIR and MIR photometric properties of (1) the most luminous 
stars in M32 and (2) the isophotal properties of the integrated light. 
The data sample wavelengths where a significant fraction of the light 
from the coolest stars is emitted. The sample of resolved stars is comprised mainly 
of objects evolving on the upper AGB, as crowding in the MIR 
data effectively prevents the detection of RGB stars (Section 4.2).
The GN observations are of particular interest for isophotal 
studies as they explore the central regions of the galaxy 
between 3 and $5\mu$m with sub-arcsec angular resolution, which is not possible 
with the current generation of space-based infrared facilities. 
Wavelength-dependent differences in isophotal characteristics such as eccentricities 
and fiducial radii could point to age gradients and/or distinct structural components.

	A distance of $770 \pm 40$ kpc for M32, measured from 
the brightness of the RGB-tip by Evans et al. (2000) and 
corresponding to a distance modulus $\mu_0 = 24.43 \pm 0.11$, 
is adopted here. Our conclusions would not change 
if the distance modulus were varied by $\sim \pm 0.1$ magnitude. A foreground reddening 
of A$_V = 0.17$, taken from the Schlafly \& Finkbeiner (2011) re-calibration of the 
Schlegel, Finkbeiner, \& Davis (1998) maps, is also used throughout this paper. 
The absence of warm and cool interstellar dust (Gordon et al. 
2006) and the colors of RR Lyraes (Sarajedini et al.  2012) 
suggests that there is no internal reddening. Finally, the `average' reddening 
law listed in Table 1 of Indebetouw et al. (2005) is adopted for MIR wavelengths. 

	Details of the observations and their reduction 
are described in Section 2, while the procedures used to construct, calibrate, and 
characterize the photometric measurements are discussed in Section 3. The CMDs obtained 
from these data are presented in Section 4, and the NIR and MIR 
luminosity functions (LFs) are compared with models in Section 5. 
The isophotal properties of M32 in the NIR and MIR 
are examined in Section 6. The paper closes in Section 7
with a brief discussion and summary of the results.

\section{OBSERVATIONS \& DATA REDUCTION}

\subsection{SPITZER IRAC Data}

	Post Basic Calibrated Dataset (PBCD) mosaics of IRAC (Fazio 
et al. 2004) images in [3.6], [4.5], [5.8], and [8.0] from 
program ID 3400 (PI Rich) were downloaded from the NASA/IPAC Infrared Science 
archive\footnote[1]{http://irsa.ipac.caltech.edu/Missions/spitzer.html}. 
The [5.8] mosaic of M32 and the surroundings is shown in Figure 1. 
The data were recorded with on-sky offsets to cover a field 
that is larger than that sampled by a single exposure, and the total 
exposure time varies across the mosaiced field. 
Only the area in the mosaics that has the largest effective 
exposure time, and hence the deepest and most uniform 
photometric characteristics, was considered for this study. The extracted M32 field 
covers a $5.0 \times 5.4$ arcmin$^2$ area that is 
centered near $\alpha$ = 00:42:56, $\delta$ = 40:45:48 (E2000).

	The images in all four IRAC channels were used to obtain the 
brightnesses of individual bright stars in and around M32, thereby 
allowing MIR CMDs and LFs to be constructed. In addition, the 
[3.6] and [4.5] images were used to examine the isophotal properties of M32.
The [5.8] and [8.0] images were not considered for this latter analysis as 
the isophotal properties of M32 could not be traced 
to radii $> 1$ arcmin in those images.

	The geometry of the IRAC detector arrays allows parallel [3.6] and [5.8] 
observations to be recorded of a field that samples the outer disk 
of M31 directly to the east of M32. This field also overlaps with moderately 
deep archival WIRCam data (Section 2.2). A sub-section of this field that covers 
a $5.1 \times 5.2$ arcmin$^2$ area and is centered near $\alpha$ = 00:42:56, 
$\delta$ = +40:45:48.3 (E2000) was used to assess contamination from 
the M31 disk in the M32 IRAC data. The location of the M31 outer disk field 
is indicated in Figure 1.

	Parallel observations that sample a field to the west 
of M32 were also recorded in [4.5] and [8.0]. This parallel field samples 
an area that is closer to the main body of M31 than M32. The stellar 
content is less homogeneous in this field because of increased contamination 
from the main disk of M31, raising the concern that it 
may not be representative of the M31 disk near M32. Therefore, this 
parallel field was not considered when assessing disk contamination.

\subsection{CFHT WIRCam Data}

	$J$ and $K$ photometry of stars in M32 and the outer 
disk of M31 were obtained from archival CFHT WIRCam (Puget et al. 2004) images. 
The detector in WIRCam is a mosiac of four $2048 \times 2048$ HgCdTe arrays 
arranged in a $2 \times 2$ format with 45 arcsec gaps between arrays. 
Each WIRCam pointing covers a $21 \times 21$ arcmin$^2$ field 
with 0.3 arcsec pixel$^{-1}$ sampling.

	The images used here are of pointings M31-401 and 
M31-405 from programs 2009BC29 (PI: Sick) and 2009BH52 (PI: Tully). 
Twenty-five 20 sec $J$ exposures and twenty-six 20 sec $Ks$ exposures 
that had been processed with the I'IWI pipeline were downloaded from 
the CFHT archive at the Canadian Astronomical Data Centre  
\footnote[2]{http://www1.cadc-ccda.hia-iha.nrc-cnrc.gc.ca/en/cfht/}. These 
were aligned to correct for on-sky offsets and the 
median of the pixel intensity distribution, which serves as a proxy for the 
mean sky level, was subtracted from each frame. The results were then stacked by 
finding the median intensity at each pixel location, trimmed 
to the area of common angular coverage, and transformed to the 
orientation of the IRAC images using the GEOMAP and GEOTRANS tasks in IRAF. 
Stars in the final images have 0.7 arcsec FWHM.

\subsection{Gemini NIRI Data}

	Ground-based observations in the MIR are hampered by high levels 
of emissivity, which mushrooms in size at wavelengths longward of $2.5\mu$m to produce 
noise of monumental proportions. The high background levels hinder efforts to 
study all but comparatively bright objects in the MIR with ground-based telescopes. 
This severe handicap notwithstanding, the large apertures of ground-based facilities 
results in angular resolutions that surpass those delivered by 
space-based telescopes, providing a niche for ground-based MIR observing. 
Ground-based observations are at present the only means of obtaining 
sub-arcsec angular resolution images in the MIR.

	The central regions of M32 were observed with NIRI (Hodapp et al. 2003) 
through $Ks$, $L'$, and $M'$ filters as part of queue program GN-2006B-Q-64 
(PI: Davidge). The NIRI detector is a $1024 \times 1024$ InSb array. 
The f/32 camera was used for these observations, and each exposure thus samples a $22.5 
\times 22.5$ arcsec$^2$ field with 0.022 arcsec pixel$^{-1}$ sampling. 

\subsubsection{$Ks$}

	The $Ks$ images were corrected for wavefront distortions using the GN 
facility AO system ALTAIR (Herriot et al. 2000). ALTAIR was used in natural 
guide star (NGS) mode, with the semi-stellar nucleus of M32 serving as the NGS. 
Twenty images, each consisting of three coadded 5 sec exposures, 
were recorded. A dither pattern that samples the corners of a $3 \times 
3$ arcsec square was used to facilitate the identification and suppression of 
cosmic rays and cosmetic defects. Five exposures were recorded at 
each dither position. The telescope was offset to one of four background fields 
that are located 90 arcsec to the North, South, East, and West of the galaxy 
nucleus after each dither position was observed. These background fields monitor 
the sky level, and allow a calibration frame to be constructed that can be used 
to remove thermal emission from warm sources along the optical path (see below).

	The reduction of the $Ks$ images followed standard procedures for NIR 
imaging, and utilized three calibration frames. The first of these 
monitors the dark current, and this was constructed by median-combining 
frames taken of the unilluminated array that have the same exposure times as 
the science data. The resulting dark frame was subtracted from each image.

	The second calibration frame monitors flat-field variations, and 
was constructed from a series of exposures that involve the Gemini 
facility calibration unit (GCAL). A series of images with the GCAL light 
source turned on were recorded, and these were followed 
by another sequence in which the GCAL light source was turned off. The first set of 
images contain signatures of the flat-field pattern (multiplicative in nature) 
combined with thermal emission (additive in nature), whereas the second set contains 
only signatures from thermal emission. Taking the difference between the two sets of 
exposures thus removes the thermal emission, leaving behind the flat-field 
information. The difference between the two sets of images was normalized to 
unity and the dark-subtracted images were divided by the resulting flat-field frame.

	A third calibration frame was constructed to remove thermal 
signatures that are produced by warm objects along the optical path, such 
as dust on exposed optical surfaces. Such objects glow in the infrared and are out 
of focus, producing diffuse signatures that usually become evident only after 
flat-fielding. Flat-fielded images of the background fields were median-combined 
after the mean sky level was subtracted from each image to remove variations 
in the sky brightness to construct a thermal contamination calibration frame. 
The result was subtracted from the flat-fielded M32 images. 

	The images that were corrected for thermal emission were aligned 
and combined by taking the median intensity at each 
pixel location. The result was trimmed to the area common to all exposures. Stars in 
the final $Ks$ image have FWHM = 0.09 arcsec. 

\subsubsection{$L'$ and $M'$}

	ALTAIR could not be used for the $L'$ and $M'$ observations 
as it has transmissive optical components that introduce background 
signal at wavelengths longward of $2.5\mu$m that saturates the NIRI 
detector. Even without ALTAIR, the high ambient background in $L'$ and $M'$ 
necessitates the use of short exposures to prevent saturation. 
A single observation at each dither position in $L'$ consisted of 
30 co-added 0.6 sec exposures, while a single observation in $M'$ consisted of 
nine exposures, each of which was made up of 34 co-added 0.5 sec exposures. 
The $L'$ and $M'$ observations were recorded with a 5 point dither pattern that 
samples the four corners and the middle of a $3 \times 3$ arcsec square. 
The dither pattern was repeated seven times in each filter, and the 
total on-source exposure times are 630 sec in $L'$ and 5355 sec in $M'$.
One of four background fields located along the cardinal axes was observed following 
each complete observation at a dither position.

	The construction of flat-field frames in $L'$ and $M'$ is 
problematic given the high levels of thermal emission in the MIR. 
In any event, thermal noise dominates in these filters, and so the 
primary processing step is the removal of thermal emission signatures. The mean 
intensity and noise pattern in the background field observations varies from 
exposure-to-exposure, with the largest variations in $M'$. 
To account for this variability, a calibration image was constructed 
for each M32 exposure by linearly interpolating between successive background field 
observations. The interpolated images were then subtracted from the M32 exposures. 
The resulting background-subtracted images have substantially lower residual 
noise than those constructed using either the preceeding or subsequent background 
observation. Even so, residual thermal signal at the few percent level remains, 
indicating that the variations in the thermal background 
vary in a manner that is more complicated than the linear trend assumed here. 

	The background-subtracted $L'$ and $M'$ frames were aligned and the median 
intensity at each pixel was found after a residual mean sky level, which was 
measured near the edges of the field, was subtracted from each image. 
The $M'$ images were aligned using the nucleus of M32, which is the only source 
detected in individual frames. The relatively modest signal from the 
center of M32 in $M'$ introduces $\leq 0.1$ arcsec (i.e. a few pixels)
uncertainties in the alignment of the individual $M'$ exposures.

	The final processing step was to trim the stacked $L'$ and $M'$ images to the 
area that is common to all pointings. Individual stars are 
seen in the final $L'$ image, and these have FWHM = 0.18 arcsec. 
As for the $M'$ images, these were recorded over two nights during only fair 
seeing conditions, and individual stars are not detected. 
The angular resolution of these data, measured from observations of standard 
stars that were observed throughout the M32 observing sequence, is 0.40 arcsec FWHM.

\section{PHOTOMETRIC MEASUREMENTS}

\subsection{SPITZER Data}

	The IRAC Instrument Handbook \footnote[3]
{http://irsa.ipac.caltech.edu/data/SPITZER/docs/irac/iracinstrumenthandbook/} 
advises against the use of point spread function (PSF)-fitting to extract photometry 
from IRAC images, with non-uniformities in pixel sensitivity and the 
undersampling of the PSF cited as concerns. Still, numerous studies of the stellar 
contents of nearby galaxies have used PSF-fitting to extract photometry from 
IRAC images (e.g. Meixner et al. 2006, McQuinn et al. 2007; Mould et al. 2008), as it 
is an accepted means of performing photometry in crowded 
environments. To assess the best means of measuring stellar
brightnesses in the current study, two sets of photometric measurements 
were made from the [3.6] and [5.8] images: (1) aperture photometry with brightnesses 
obtained from the PHOT routine in the DAOPHOT (Stetson 1987) package, and (2) 
PSF-fitting photometry using ALLSTAR (Stetson \& Harris 1988). 
Sources within 90 arcsec of the center of M32 were not 
photometered, as surface brightness fluctuations in this high-density region 
can masquerade as individual stars. Both sets of photometry were calibrated using 
flux zeropoints calculated by Reach et al. (2005).

	Crowding is an obvious concern when conducting aperture photometry in 
external galaxies. To mitigate against crowding, the aperture measurements were made 
with a comparatively narrow 3 pixel (1.8 arcsec) radius. Final 
brightnesses were then obtained by applying an aperture correction that was 
measured from stars in uncrowded parts of the images. An additional concern 
for crowded extragalactic environments is that non-uniformities in the integrated 
surface brightness can occur over angular scales that are comparable to the area 
where the local sky level is measured, and these introduce errors into 
both aperture and PSF-fitting measurements. To correct for localized 
sky structure, a running boxcar median filter with dimensions that are sufficient to 
suppress stars was applied to the images, and the resulting smoothed image 
was subtracted from the data.

	PSFs were constructed using tasks in DAOPHOT (Stetson 1987). 
A single PSF was constructed for each filter, and faint objects close to 
PSF stars were subtracted out using progressively improved versions of the PSF. 
Once a final PSF was obtained then the final photometric measurements were 
made with the PSF-fitting routine in ALLSTAR (Stetson \& Harris 1988). 

	The PHOT and ALLSTAR routines estimate the uncertainty in photometric 
measurements using the properties of the source and the surrounding background. 
The photometric measurements of objects that are in crowded environments and/or 
that are non-stellar in appearance have higher uncertainties than those in 
uncrowded environments that have a star-like light profile. 
To remove sources with problematic photometry, objects that 
depart from the general trend on the error $vs.$ magnitude relation for each filter 
were removed from the photometric catalogue. This parallels the procedure 
applied by Davidge (2010) to PSF-fitting photometry, where objects were rejected 
using the uncertainty computed by ALLSTAR. Davidge (2013) found that the properties 
of the final sample culled using photometric error are similar to those that 
result when other object characteristics, such as image sharpness, are 
used to identify sources with problematic photometry. 

	The $([5.8], [3.6] - [5.8])$ CMDs obtained from aperture and PSF-fitting 
measurements for sources in M32 and the M31 disk field are compared in Figure 2. 
The standard deviation about the mean [3.6]--[5.8] 
for objects with [5.8] between 15.75 and 16.25, $\sigma_{AGB}$, 
is shown for each CMD. This brightness interval samples AGB stars in a 
heavily populated part of each CMD, and is well above the faint limit. An 
iterative $2.5\sigma$ rejection filter was applied to suppress outliers when 
computing $\sigma_{AGB}$.

	The CMDs constructed with PSF-fitting sample a larger number of 
objects and go significantly fainter than those obtained from 
aperture measurements. While the differences between $\sigma_{AGB}$ 
computed from the aperture and PSF-fitting photometric measurements are modest, 
the AGB sequence when [5.8] $< 15$ in the PSF-fitting CMD of M32 
is better defined than in the aperture photometry CMD. Given (1) the 
differences in the number of objects between the two sets of CMDs, and (2) the 
tightness of the bright end of the M32 CMD obtained with PSF-fitting, 
then PSF-fitting is adopted as the prefered photometry technique.

\subsection{WIRCam and Gemini NIRI Data}

	The PSF-fitting program ALLSTAR was used to obtain photometric measurements 
from the WIRCam and NIRI images. As with the IRAC observations, the PSFs were 
constructed from bright, isolated stars, and an iterative scheme was applied to suppress 
faint neighboring stars. No attempt was made to photometer objects within 90 
arcsec of the galaxy center in the WIRCam images, as crowding confounds efforts to 
resolve individual stars there unless the data have an angular resolution 
$\lesssim 0.1$ arcsec FWHM (e.g. Davidge et al. 2000). 
The WIRCam photometry was calibrated using zeropoints obtained from 
standard star observations that were recorded during the same semester, and 
this information is available on the CFHT web site \footnote[4]
{http://www.cfht.hawaii.edu/Instruments/Imaging/WIRCam/WIRCamStandardstars.html}. 
The NIRI photometry was calibrated using zeropoints obtained from standard star 
observations that were recorded as part of the baseline calibration set for 
this program.

	The IRAC observations have angular resolutions that are significantly 
poorer than the WIRCam data. The WIRCam data can then be used to assess in 
a preliminary way the effect of crowding on the MIR data. Emphasis is placed on the 
[3.6] data for these experiments, as (1) this passband is closest in wavelength 
to the WIRCam observations, and (2) at longer wavelengths crowding is expected to 
become less of an issue because of the increased contrast between the brightest AGB 
stars and the vast majority of other stars in M32.

	Stars in the IRAC [3.6] observations have a FWHM of 1.8 arcsec, and 
so the $J$ and $Ks$ WIRCam images were convolved with a Gaussian that degrades 
the angular resolution to match that of the [3.6] observations. 
The $(K, J-K)$ CMDs of objects between 1.5 and 2.5 arcmin distance from the 
center of M32 obtained from the unsmoothed and smoothed WIRCam images are compared in 
Figure 3. While there are obvious differences near the faint end, there is reasonable 
agreement at the bright end, most noteably near the AGB-tip. 

	The contrast between AGB stars and the vast body of 
fainter stars increases towards longer wavelengths, and so 
the method described above is a conservative approach for assessing 
the impact of crowding. It is thus significant that the comparisons in 
Figure 3 suggest that the angular resolution of the [3.6] data should not skew 
the properties of the brightest stars in the outer regions of M32. Additional evidence 
to support this claim is presented in Section 4.2.

\section{RESULTS: CMDs}

\subsection{Near-Infrared CMDs}

	The $(K, J-K)$ CMDs of the M31 outer disk and 
of three annuli centered on M32 that sample equal areas on the sky 
are shown in Figure 4. The inner and outer boundaries of the region where 
resolved stars in M32 are investigated are indicated in Figure 1.
The dominant plume of objects in each CMD is populated by 
AGB stars, and the $J-K$ color and peak brightness of the AGB sequence are 
consistent with those measured by Davidge \& Jensen (2007). 
It should be recalled that the WIRCam CMDs sample a much larger portion of M32 than 
those presented by Davidge \& Jensen (2007). The peak $K$ brightness in the Inner 
annulus is $K \sim 15.5$, which is similar to the peak brightness near 
the center of the galaxy (e.g. Davidge et al. 2000). The absence of a 
radial trend in the peak $K$ brightness is one indicator that the 
photometry of the brightest stars is not affected by crowding.
The M31 CMD extends to fainter magnitudes than the M32 CMDs, owing 
to the lower stellar density in that field. 

	A significant fraction of the brightest AGB stars near the center of M32 are 
long period variables (LPVs) with light curve amplitudes in $K$ that may exceed two 
magnitudes (Davidge \& Rigaut 2004). Such variability complicates efforts to determine 
ages from CMDs using the brightest stars. To mitigate the impact of such variability, in 
the remainder of this section emphasis is placed on the photometric properties of stars 
that are midway down the AGB on the CMD, where the affects of 
LPV-like variations should average out. The SFH of M32 is investigated 
in Section 5 using model LFs that account for LPV behaviour. 

	Basic insights into the stellar contents of M32 and the outer disk of M31 can 
be gleaned by comparing the mean colors of the AGB sequences. To this end, mean 
$J-K$ colors were computed for objects with $K$ between 16.75 and 17.25. This 
magnitude interval samples an area of the CMD that is well above the faint limit and is 
richly populated in both M32 and M31. A $2.5\sigma$ 
iterative rejection filter was applied to suppress outliers.

	The mean $J-K$ colors in the CMDs of the Inner, Middle, and Outer annuli 
of M32 are $1.255 \pm 0.010$, $1.277 \pm 0.011$, and $1.292 \pm 0.015$, 
where the uncertainties are the standard error of the mean. 
Modelling of long-slit spectra suggests that the 
stellar content of M32 changes with radius (Worthey 2004; Rose et al. 2005; Coelho et 
al. 2009). Still, the systematic absorption line gradients that are typical of 
large pressure-supported galaxies (e.g. Davidge 1992; Davies et al. 
1993; Sanchez-Blazquez et al. 2006) are not seen 
(Davidge 1991; Davidge et al. 2008, but see also Hardy et al. 1994). 
Some of the features that do vary with radius in M32 show behaviour that is 
contrary to what might be expected when compared with classical elliptical galaxies 
(e.g. Davidge, de Robertis \& Yee 1990).

	The progressive increase in $J-K$ towards larger radius in M32 
is contrary to what is expected in the absence of large-scale spectroscopic 
gradients. Rather than sampling an instrinsic radial change in the properties of M32, 
it is likely that the radial increase in mean $J-K$ is due to contamination from 
stars in the M31 disk. The mean $J-K$ color in the M31 disk field is $1.343 \pm 0.010$, 
and the fractional contamination from M31 disk stars in the M32 CMDs increases towards 
larger radii. Indeed, $K-$band surface brightness measurements of M32 from the 2MASS 
Large Galaxy Atlas (Jarrett et al. 2003) indicate that the number of stars in the 
Outer annulus that belong to M32 is 75\% lower than in the Inner annulus. However, the 
Outer annulus contains 40\% fewer stars than in the Inner annulus, indicating that 
the fractional contribution from M31 disk stars grows with increasing radius. 
Adding an AGB component with a color like that in the M31 disk field to a 
population with a mean color like that in the Inner annulus will produce a 
radial trend in the mean M32 $J-K$ colors in the same sense as that found here.

	The mean $J-K$ color of AGB stars in M32 is bluer on average than in the 
M31 disk, suggesting differences in mean metallicity and/or mean age. 
Quantitative insights into the stellar contents of M32 and the M31 
outer disk can be obtained by making comparisons with isochrones, and 
the CMDs of the M31 outer disk and the M32 Middle annulus are compared 
with model sequences from Marigo et al. (2008) in Figure 5. These 
sequences, and all other models used in this paper, were 
downloaded from the Padova Observatory website \footnote[5]
{http://stev.oapd.inaf.it/cgi-bin/cmd}, and are based on the most recent versions 
of the models constructed by this group that include the thermally-pulsing (TP) 
phase of AGB evolution. This phase of evolution is critical for understanding the 
MIR properties of the most evolved AGB stars.

	The models shown in Figure 5 have Z = 0.016 and 
ages 1 Gyr, 3 Gyr, and 10 Gyr. Based on previous estimates of the metallicity of 
M32, it is anticipated that the number of C stars will be modest, and 
the majority of stars will be oxygen-rich M giants. Thus, the 
60\% AlOx $+$ 40\% silicate composition model discussed by Groenewegen (2006) 
is assumed for circumstellar dust. While the chemical composition of the circumstellar 
disk is of only secondary importance for the NIR photometric properties of AGB stars, 
this is not the case in the MIR (Section 4.2).

	The 3 Gyr isochrone tracks the ridgeline of the M32 sequence in the right hand 
panel of Figure 5, while the 1 and 10 Gyr isochones more-or-less follow the 
blue and red envelopes of the AGB sequence. In contrast, the majority of bright stars 
in the M31 outer disk have isochrone-based ages $\geq 3$ Gyr when compared with the 
solar metallicity sequences. That the AGB stars in the M31 field tend to have colors 
that are suggestive of a sizeable old population is consistent with the SFH measured in 
other fields in the outer disk of M31 (e.g. Davidge et al. 2012; Bernard et al. 2012), 
including close to M32 (Monachesi et al. 2012). 

	A fiducial Galactic RSG sequence, constructed from the entries in Table 5 
of Koornneef (1983), is included in the left hand panel of Figure 5. 
The objects that populate the plume with $J-K \sim 0.6$ and 
$K < 16$ in the M31 outer disk CMD have colors that are consistent 
with them being G and K supergiants, indicating that the outer disk 
field contains moderately young stars. Relatively young stars are expected 
in this part of M31 as (1) there are areas of recent star formation 
in nearby fields (van den Bergh 1964; Hodge 1979), and (2) members of other 
groupings may drift into this field due to the motions that 
these objects attain as they interact with large interstellar 
clouds. Significant mixing of stars in the low-density outer regions of disks 
can occur over time scales of only a few tens of Myr (e.g. Davidge et al. 2011).

\subsection{Mid-Infrared CMDs}

	The $([5.8],[3.6]-[5.8])$ CMDs of objects in M32 and the M31 outer disk 
are shown in Figure 6, while the $([8.0], [4.5]-[8.0])$ CMDs of M32 are 
shown in Figure 7. The majority of stars in these CMDs are evolving on the AGB, 
and there are few -- if any -- RGB stars. The Marigo et al. (2008) 
models predict that the red giant branch (RGB)-tip in solar metallicity 
systems with ages $> 3$ Gyr has an absolute magnitude $\geq -7$ in each IRAC filter, 
which corresponds to apparent magnitudes $\geq 17.5$ in M32. 
Given that the number density of RGB stars is $\sim 4 - 5 \times$ 
higher than that of AGB stars, then it is perhaps not unexpected that 
the faint limit of the $([5.8],[3.6]-[5.8])$ CMDs coincides with the expected 
location of the RGB-tip, as the resulting increase in crowding near the RGB-tip 
will impose a limiting magnitude if individual RGB stars can not be resolved.

	The vertical sequence with [3.6] -- [5.8] $\sim 0$ in the left 
hand panel of Figure 6 contains a mix of foreground main sequence stars, evolved 
stars in M31 and M32 that do not have circumstellar envelopes, and extragalactic 
sources. This plume is less well populated in the M32 $([5.8], [3.6]-[5.8])$ CMDs 
due to the smaller sky coverage of each annulus. The number of 
extragalactic sources in the CMDs can be estimated from 
the source counts given by Fazio et al. (2004). Assuming a 
uniform distribution on the sky then there should be $\sim 74$ extragalactic 
objects in the M31 outer disk field with [5.8] between 15.5 
and 16.5, and 56 objects towards M32. For comparison, there are 359 objects detected
in this magnitude range in the M31 outer disk field, and 432 
towards M32. Background galaxies thus account for 10 -- 20\% of the objects in 
the $([5.8],[3.6]-[5.8])$ CMDs. These are upper limits to extragalactic 
contamination as ALLSTAR rejects sources that are obviously non-stellar.

	The M31 outer disk field contains objects with [5.8] $= 13.5 - 14.0$ that 
span a broader range of $[3.6]-[5.8]$ colors than those in the M32 CMDs. 
There is evidence that many of the brightest objects in the M32 Middle and Outer 
annulus CMDs likely belong to the M31 disk. In particular, outside of the region within 
90 arcsec of the center of M32 where individual stars are most susceptible to 
blending, objects with $[5.8] \geq 14$ appear to have a more-or-less uniform 
distribution, as expected if they belong to the outer disk of M31.

	The $([5.8],[3.6]-[5.8])$ and $([8.0],[4.5]-[8.0]$) CMDs of the Middle annulus 
of M32 are compared with Z=0.016 isochrones from Marigo et al. (2008) in Figure 8. 
While the 1 and 3 Gyr isochrones match the ridgeline of the M32 $([5.8],[3.6]-[5.8])$ 
CMD, the agreement with the brightest objects is poor. 
The isochrones on the $([8.0], [4.5]-[8.0])$ CMD also fall short of the 
brightest stars in the Middle annulus of M32, although 
the ridgeline of $[4.5]-[8.0]$ colors is consistent with ages $\sim 3$ Gyr.
The inability of the models to match the observations near the bright end may be 
because some of the brightest AGB stars are LPVs. Comparisons with the 
$([3.6]-[4.5],[4.5]-[5.8])$ two color diagram (TCD) -- discussed below -- indicates that 
the predicted spectral-energy distribution (SEDs) disagree with the observations 
by 0.1 -- 0.2 magnitudes.

	The isochrones also indicate that old, metal-rich AGB stars 
should be detected with these data. The 10 Gyr isochrone more-or-less skirts 
the red envelope of objects in M32 when $[5.8] < 17$, with LPVs possibly accounting for 
the modest number of objects with [3.6]--[5.8] colors that are redder than the 10 Gyr 
sequence. A similar situation holds for the $([8.0], [4.5]-[8.0])$ CMD. There is thus 
photometric evidence in the MIR for an old metal-rich population, and in Section 6 
it is demonstrated that these objects contribute significantly to the number counts.

	The $([5.8], [3.6]-[5.8])$ CMD of the M31 outer disk field is compared 
with isochrones in the left hand panel of Figure 8. The 
main locus of the $([5.8], [3.6]-[5.8])$ CMD of the M31 
outer disk when $[5.8] < 16$ is consistent with a mean age that is 
similar to M32. The isochrones suggest that the 
objects with [3.6]--[5.8] $> 1$ near [5.8] $\sim 14$ 
may be highly evolved AGB stars with ages between 3 and 10 Gyr. Still, some 
of these may be LPVs, in which case their location on the CMD may not reflect 
their true age and metallicity. The $([5.8], [3.6]-[5.8])$ CMD of the M31 outer disk 
field is consistent with stars being present that formed over a broad range of ages.

	TCDs are a means of comparing model SEDs with those that are observed. 
The $([3.6]-[4.5],[4.5]-[5.8])$ TCDs of sources in M32 with [4.5] $< 17$ are 
compared with post-RGB models from Marigo et al. (2008) in Figure 9. There is a 
red plume in each radial interval that has a dispersion of a few tenths of 
a magnitude. Even though the stars on this plume are highly evolved and 
many expected to be LPVs, the comparatively modest scatter is likely a consequence of 
the simultaneous nature of the IRAC observations, coupled with the 
smaller amplitude of LPV variations towards longer wavelengths. Still, the red 
plumes predicted by the models systematically fall 0.1 -- 0.2 magnitudes below 
the observations. 

	The properties of isochrones on the upper reaches of MIR CMDs depends on 
the composition of the circumstellar dust, as this material can be a significant 
source of emission at these wavelengths. The comparisons in Figures 8 and 9 assume 
that circumstellar dust around M stars has a 60\% ALOx $+ 40$\% silicate mix. 
Do the ages obtained above change if different dust compositions are adopted? The 
role that dust composition plays on MIR isochrones is investigated 
in Figure 10, where the $([5.8],[3.6]-[5.8])$ and $([8.0], [4.5]-[8.0])$ 
CMDs of the M32 Middle annulus are compared with Z=0.016 isochrones from Marigo 
et al. (2008) that assume 100\% silicate (solid lines) and 
100\% ALOx (dashed lines) compositions. 

	At a fixed age, the comparisons in Figure 10 indicate that model-to-model 
differences in peak brightness amount to only a few tenths of a magnitude. 
The mean [3.6]--[5.8] color of the 3 Gyr isochrone is also not 
sensitive to the chemical properties of the dust, although this 
is not the case for the 10 Gyr [3.6]--[5.8] colors. 
Dust composition has a greater impact on isochrones in the $([8.0], 
[4.5]-[8.0])$ CMD, due to the greater contribution from thermal emission near [8.0] 
and the differences in absorption efficiencies that grow with 
wavelength (e.g. Figure 1 of Groenewegen 2006). Indeed, the 3 
and 10 Gyr isochrones that assume a 100\% ALOx composition overlap on 
the $([8.0], [4.5]-[8.0])$ CMD. Neither 100\% ALOx sequence gives a satisfactory match 
to the M32 CMD, with both falling redward of the M32 ridgeline. This suggests 
that some silicates must be present to reproduce the [4.5]--[8.0] colors of evolved AGB 
stars in M32. 

\subsection{The $([3.6], K-[3.6])$ CMD}

	The NIRI $K'$ and $L'$ images are of special interest as they have an 
angular resolution that allows the brightest stars to be resolved to 
within a few arcsec of the galaxy nucleus. The $L'$ magnitudes can 
be transformed into the IRAC [3.6] photometric system, allowing the behaviour of 
objects on the $([3.6], K-[3.6])$ CMD to be examined over a wider range 
of radii than is possible with the IRAC data alone. The 
$([3.6],K-[3.6])$ CMDs of stars in M32 and the M31 outer disk are shown in Figure 11. 
The M32 Center CMD is constructed exclusively from the NIRI observations. Objects 
within 2 arcsec of the nucleus of M32 have been excluded, as the 
photometry in that region is compromised by crowding and/or surface brightness 
fluctuations. The CMDs in the other panels use $K$ magnitudes from 
the WIRCam observations and [3.6] magnitudes from the IRAC images. 

	The peak [3.6] magnitudes in the M32 Center, Inner, and Middle annuli agree to 
within $\pm 0.2$ magnitude. Thus, the AGB peak brightness does not change 
greatly with radius, and there is not a central concentration of luminous AGB stars. 
With the exception of the outer annulus CMD, the peak [3.6] brightness of objects 
in the M31 outer disk field is $\sim 0.5$ magnitude brighter than in the M32 CMDs. 
If the brightest stars in the M31 outer disk were uniformly distributed then 
similar objects might be expected in the inner and middle annuli, although 
a population of bright field stars would not be expected in the M32 Center field 
due to the modest area covered on the sky when NIRI is used in f/32 mode. 

\section{RESULTS: LFs AND THE SFH DURING INTERMEDIATE EPOCHS}

	The LFs of stars in the inner, middle, and outer annuli of M32 are compared 
in Figure 12. Contamination from foreground stars, background galaxies, and stars in 
the disk of M31 is a concern, and so the M31 outer disk LFs were 
subtracted from the M32 LFs after adjusting for differences 
in sky coverage. A potential source of uncertainty in this 
correction is that the stellar content of the M31 disk field 
may differ from that of the disk near M32 (e.g. Figure 11).

	The entries in Figure 12 are specific frequencies that 
give the number of objects per 0.2 magnitude interval in a system with M$_K = -16$. 
The scaling to M$_K = 16$ was done using integrated annular brightnesses computed 
from the M32 surface brightness measurements in the 
2MASS Large Galaxy Atlas (Jarrett et al. 2003). It can be seen from 
Figure 12 that the specific frequency in $K$ does not vary significantly 
from annulus to annulus, in agreement with the results found by Davidge \& 
Jensen (2007). However, there is a tendency for the specific frequencies 
of stars in the [5.8] LF of the outer annulus to fall below those 
in the inner and middle annuli, although the differences for individual bins are not 
statistically significant. It should be recalled that the $K$ and [5.8] passbands 
sample different sources of emission, with the [5.8] LFs being more sensitive to 
AGB stars that have circumstellar envelopes than the $K$ LFs. 

	The interpretation of the LF of AGB stars is complicated by the rapid pace 
of evolution on the upper AGB, so that small number statistics may make 
signatures associated with changes in the SFR difficult to detect. 
Photometric variablity will also blur the LFs, smoothing features that 
might otherwise provide insight into the SFH. Finally, models of AGB 
evolution are prone to uncertainties in the stellar physics (e.g. Marigo et al. 2008). 
The expectation of uncertainties in the models notwithstanding, it is encouraging that 
there is consistency between the isochrones and observations on the 
CMDs and TCDs (Section 4). 

	Rather than attempt to deduce an independent SFH from the LFs, the present
analysis is restricted to comparisons with models that follow pre-defined 
SFHs. The simplest models are those that assume a simple stellar population (SSP), 
and in the top panel of Figure 13 the mean $K$ LFs of AGB stars in the inner and middle 
annuli are compared with 2 and 10 Gyr SSP sequences constructed from the Marigo et al. 
(2008) models. The models assume Z=0.016, a 60\% silicate $+$ 40\% AlOx 
circumstellar dust mix, and that none of the stars are photometric variables. 
The models in this panel have been scaled to match 
the number counts in the two faintest bins. 

	The 10 Gyr model covers only the faintest part of the plotted LF. 
In contrast, the 2 Gyr model in the top panel of Figure 13 provides a 
reasonable match over many magnitudes to the $K-$band LF. Still, 
the 2 Gyr model predicts a peak $K-$band magnitude that is $\sim 1$ magnitude 
fainter than observed. Comparisons involving the [5.8] LF yield 
similar results.

	The majority of bright AGB stars in M32 are LPVs, with light 
curve amplitudes that approach $\pm 1$ magnitude in $K$ (Davidge \& Rigaut 2004), and 
the light distribution in the models changes significantly if LPV-like photometric 
variability is added to the models. This is demonstrated in the middle panel 
of Figure 13, which shows the 2 Gyr model $K$ LF convolved with the $\Delta K$ 
distribution that Davidge \& Rigaut (2004) obtained from 
the light curves of Galactic bulge LPVs tabulated by Glass et al. (1995). 
After convolution with this amplitude kernel, the 2 Gyr model $K$ LF matches the 
peak of the observed LF as well as the approximate shape of the LF near the bright end. 

	Photometric variability can also be folded into models that 
cover the IRAC passbands. The amplitude of LPVs near $5\mu$m is $\sim 0.8 
\times$ that in $K$ (e.g. Table 4 of Le Bertre 1993), and so 
a modified $K$ kernel, that was compressed along the magnitude axis by a 
factor $1/0.8 = 1.25 \times$, is applied to the [5.8] models. 
The 2 Gyr model constructed in this manner is shown 
in the bottom panel of Figure 13, and there is reasonable agreement with the [5.8] LF.

	While the comparisons with the 2 Gyr SSP model in Figure 13 are consistent 
with a large intermediate age population, in reality M32 
is a composite stellar system containing stars that span a range of ages 
and metallicities. Monachesi et al. (2012) have investigated the SFH of M32 
using stars near the MSTO, and model LFs that are based on the 
SFH shown in their Figure 13a are compared with the observations in the middle 
and lower panels of Figure 13. The models follow a Chabrier (2001) mass function, 
and were convolved with the $K$ and [5.8] amplitude kernels described above. 
Tacit assumptions are that: (1) all AGB stars regardless of mass have the same 
variability characteristics, and (2) all of the AGB stars in M32 are LPVs. 
The models further assume that the brightest AGB stars have a solar 
metallicity, which is not unreasonable given the metallicity distribution 
function and age-metallicity relation found by Monachesi et al. (2012).

	The composite model matches the $K$ number counts when $K > 15.8$, and 
is a better match to the observations in this magnitude range than the 2 Gyr SSP LF. 
The improved agreement with the observations is because old AGB stars in the 
composite model bolster the number counts at the faint end, making the composite 
model steeper than the SSP model. The agreement between 
the composite model and observations degrades when $K < 15.7$, although the 
differences are at less than the $2\sigma$ level.
The agreement with the [5.8] LF is similarly good when 
[5.8] $> 14.4$, with the overall shape of the observed LF reproduced 
at these magnitudes. The agreement is also better than 
was achieved with the 2 Gyr SSP model. However, as with the $K$ LF, the 
composite model does not match the bright end of the LF as well as the 2 Gyr SSP model.

	Despite the promising agreement between the composite model and the 
observations, the model falls short near the bright ends of the $K$ and [5.8] LFs. 
There are uncertainties in the structure models of 
the most evolved AGB stars, and the cumulative impact 
of any uncertainties are expected to be greatest near the AGB-tip. 
Uncertainties in the LPV amplitude distribution function may also 
contribute to difficulties matching the numbers of very bright objects. 
A single amplitude kernel that is based on one population of objects (LPVs in 
the Galactic bulge) was adopted for this study. The agreement between the 
models and the bright end of the LFs in Figure 13 would improve if the amplitude 
distribution function were to be extended to larger amplitudes and/or was skewed 
to contain a larger fraction of objects at the high amplitude end.

	The differences between the observations and the 
composite models might also indicate a mix of ages that differs from that predicted by 
the Monachesi et al. (2012) SFH. A larger fractional contribution from moderately 
young stars should boost the number of stars near the observed AGB-tip. To investigate 
this possibility, a model LF that follows the Monachesi et al. 
(2012) SFH from the present day to 7 Gyr, but that has no stars older than 7 Gyr, 
was constructed and the result is shown in Figure 13.

	The suppression of star formation more than 7 Gyr 
in the past has a significant impact on the agreement with the observations. 
The differences are more pronounced in [5.8] than in $K$ as the AGB sequences of old 
populations extend to brighter intrinsic magnitudes in the MIR. 
The comparison between the [5.8] LF and the truncated SFH model thus indicates 
that M32 must contain a substantial metal-rich population that formed 
during relatively early epochs. This demonstrates that MIR observations provide a 
direct means of probing the SFH of a system over a wide range of epochs. 

	How sensitive is the [5.8] LF to changes in the 
size of star-forming events during intermediate epochs? 
The [5.8] LF of M32 is compared with models that contain an old population 
and an intermediate age component in Figure 14. 
Two flavors of intermediate age component are investigated: a discreet 
burst, which is represented by a 2 Gyr SSP, and 
a star-forming event with a constant SFR that has a 1 Gyr duration, and 
occured between 2 and 3 Gyr in the past. A solar metallicity is assumed.

	Models in which the intermediate age component contributes 
three different stellar mass fractions are shown in Figure 14. 
It can be seen that (1) the composite models provide a 
better fit to the LFs than SSPs, and (2) models that involve an extended 
period of star-forming activity during intermediate epochs give a better match to the 
LF in the [5.8] interval between 14.5 and 16.0. In fact, the slope of the LF in this 
brightness interval, which corresponds to M$_{[5.8]}$ between --8.5 and 
--10.0, is sensitive to the age mix of the system. Models that 
assume a dominant intermediate age component yield a better match to the bright 
end of the [5.8] LF, although this is where the evolutionary models are 
least reliable and there is susceptibility to uncertainties in the nature of 
the amplitude distributions of LPVs (see above).

	To summarize this section, model LFs that assume that the 
brightest AGB stars in M32 are LPVs better match the 
LFs in $K$ and [5.8] than those that assume no variability. Models that contain an 
intermediate age component -- including those 
based on the SFH constructed by Monachesi et al. (2012) -- match the shape 
of the LFs in M32 in $K$ and [5.8] over a range of magnitudes. 
It is also shown that the presence of an old population 
is required to explain the overall shape of the [5.8] LF of M32. These results highlight 
the potential utility of MIR observations of AGB stars to probe the stellar contents 
of moderately distant galaxies.

\section{RESULTS: SURFACE BRIGHTNESS PROFILES AND STAR COUNTS}

\subsection{The Main Body of M32}

	The light profile of a galaxy forms part of its fossil record 
and can be used to gain insights into its evolution. In this section, the structural 
characteristics of M32 are examined over a range of wavelengths to search for 
possible radial differences in stellar content that might 
provide clues into the past history of the galaxy. 
To this end, the structure of M32 in $J, K, [3.6]$, and [4.5] was investigated using 
the isophote-fitting task {\it ellipse} (Jedrzejewski 1987), as implemented in STSDAS. 
The analysis was performed on star-subtracted images. In addition to 
removing resolved stars that belong to M32, star subtraction also removes bright 
stars that belong to M31, but does not account for unresolved light from 
M31. To account for this, residual background sky levels were measured near the 
edges of the observed field and the result was subtracted prior to the 
isophotal analysis. This procedure assumes that the light from the 
M31 disk is uniformly distributed across the field, with the areas where the 
background light levels are measured being representative of this part of M31. 

	The light profiles are compared in the top panel of Figure 15. 
A cluster of bad pixels causes a gap in the radial coverage 
of the WIRCam observations, and the affected interval 
is indicated. The surface brightness profile in the central 150 arcsec of M32
can be fit with an R$^{1/4}$ law (e.g. Choi et al. 2002). The effective radius, R$_e$, 
and surface brightness, $\mu_e$, obtained by fitting an R$^{1/4}$ law to the WIRCam and 
SPITZER images are listed in Table 1. The R$_e$ measurements made by Kent (1987) and 
Choi et al. (2002) at visible/red wavelengths are also listed for comparison. 

	The errors in R$_e$ are $1\sigma$ random uncertainties that reflect 
the scatter about the fitted relation. Systematic effects, which are 
not included in the error estimates, have the potential to introduce uncertainties 
that are much larger than the random errors. For example, 
Choi et al. (2002) and Kent (1987) measure R$_e$ 
at similar wavelengths, but their results differ by 3 arcsec. This 
suggests that a more realistic total uncertainty in R$_e$ may be 
$1 - 2$ arcsec, which is $2 - 3 \times$ larger than the random uncertainties.
Still, it is encouraging that the R$_e$ values measured in $K$, [3.6], and [4.5] agree 
within their random uncertainties, while the [3.6] and [4.5] R$_e$ measurements agree 
with the R$_e$ measured in the $R$ filter by Kent (1987) at the $2\sigma$ level.
Given the uncertainties in R$_e$, the entries in Table 1 suggest that R$_e$ is constant 
with wavelength, at least to within a few arcsec. The spatial distribution of the 
components that dominate the light at visible wavelengths, which includes a substantial 
contribution from main sequence stars, is thus not greatly different from the 
spatial distribution of the red stars that contribute significantly to the 
unresolved NIR and MIR light.

	Additional insights into the evolution of M32 can be gleaned from isophote 
shapes. The ellipticities measured from the WIRCam and Spitzer images are compared 
with ellipticity measurements from Table III of Kent (1987) in 
the middle panel of Figure 15. The error bars show $1\sigma$ uncertainties that 
reflect the scatter about the fitted ellipses. These are 
internal errors, and the point-to-point jitter in the eccentricity measurements in a 
given filter suggests that at large radii the actual uncertainties may be a few 
times larger than those computed by {\it ellipse}. 

	There are no systematic differences between the NIR, MIR and 
R-band ellipticities at radii $< 16$ arcsec. However, between 
16 and 45 arcsec the R-band ellipticities consistently fall 
below those measured at longer wavelengths, suggesting that the NIR and MIR isophotes 
at these radii are flatter than at visible wavelengths. The uncertainties in the 
ellipticities computed by {\it ellipse} would have to be in error by almost 
an order of magnitude to overlap with the $R$ eccentricity 
measurements. In fact, the ellipticities measured 
in $J$ and $K$ from the WIRCam data agree with those obtained by Peletier 
(1993) in the same filters, while the ellipticities measured by Kent (1987) agree 
with those measured by Choi et al. (2002). Hence, the differences found here 
are consistent with other independent measurements. 

	The ellipticity measurements suggest that at least some of the 
unresolved NIR and MIR light originates from objects that have a flatter distribution 
on the sky than the objects that dominate the light at shorter wavelengths. 
Such a distribution is suggestive of a disk, and evidence of a disk in the 
outermost regions of M32 has been found by Choi et al. (2002) and Graham (2002). 
The coefficient of the fourth order cosine term in the fourier 
expansion of the isophotes is a sensitive indicator of isophote 
shape (e.g. Carter 1978). If this coefficient $< 0$ then the isophote 
is `boxy, whereas if it is $> 0$ then the isophote is 
`disky'. The fourth order cosine coefficients -- B4 -- 
measured from the NIR and MIR data are plotted in the lower panel of Figure 15. The 
B4 values typically have only modest departures from zero, and to suppress 
random errors the points that are plotted in the lower panel of Figure 15 
are the averages of coefficients from three adjacent isophotes. 

	The entries in the lower panel of Figure 15 indicate that when 
$r < 16$ arcsec the B4 coefficient tends to be predominantly 
negative, indicating a boxy morphology. However, at larger radii there is a 
higher incidence of positive B4 values. The mean B4 values at $r < 16$ arcsec 
and at $r > 48$ arcsec are compared in Table 2, and it is evident that B4 changes with 
radius in $J$, [3.6], and [4.5]. In fact, (1) the mean B4 entries in these filters 
in a given radial interval all agree, and (2) at $r > 48$ 
arcsec the B4 coefficient in these filters exceeds zero at the $2\sigma$ 
or greater level. Still, the B4 measurements made from the 
$Ks$ images differ from those in the other three filters. Those measurements 
notwithstanding, the B4 values from the other three filters provide 
tantalizing evidence that the structural properties of M32 
when $r < 16$ arcsec is different from that at $r > 48$ arcsec.

\subsection{The Central Regions of M32}

	The center of M32 samples the deepest part of the gravitational potential 
well, and so may harbor signatures of past events that shaped its evolution. 
The angular resolution of the NIRI $K', L'$ and $M'$ observations 
allow the light distribution near the center of M32 
to be investigated on spatial scales $\leq 1$ parsec. 
The central regions of M32 have been examined previously at sub-arcsec 
resolutions, although not at wavelengths $> 2.5\mu$m. 
Lauer et al. (1998) and Corbin et al. (2001) find no evidence of radial 
color variations in the visible and NIR, while Peletier (1993) and Davidge (2000) 
find that the narrow-band CO index in M32 may strengthen with decreasing radius in the 
central 3 arcsec. 

	The $K'$ and $L'$ NIRI observations were smoothed to the 
angular resolution of the $M'$ data, and the resulting light, color, 
and ellipticity profiles are shown in Figure 16. 
The central structural properties of M32 change substantially over 
small angular scales, and subtle variations in the character of the outer 
wings of the PSF may significantly affect colors in these regions. Therefore, 
no effort is made to examine the isophotal properties of the central 0.2 arcsec of M32. 

	The light profiles in the top panel of Figure 16 have similar slopes, 
and the $K'-L'$ and $K'-M'$ colors are more-or-less constant. The 
ellipticities measured from the $K'$ and $L'$ images are similar out to 1.3 
arcsec. However, between 0.5 and 0.7 arcsec the $M'$ ellipticities fall below those in 
$K'$ and $L'$. While the centroiding of the individual $M'$ images was subject to 
$\pm 0.1$ arcsec jitter (Section 2), this can not explain the dispersion in 
ellipticities at these radii. Uncertainties due to centroiding errors were assessed by 
conducting simulations involving a mock $M'$ image that was constructed using 
the isohophotal analysis of the actual $M'$ image. As the final $M'$ science 
image was the result of median-combining 35 individual images, 
this mock image was replicated 35 times, with each replicant 
randomly offset by $\pm 0.1$ arcsec in radius. 
The 35 images with simulated image jitter were then median-combined, and 
$ellipse$ was run on the result. A number of such realizations were 
run, and a comparison of the ellipticities measured from all of these 
indicates that a $\pm 0.1$ arcsec centroiding error introduces 
$\pm 0.03$ error in the ellipticities at a radius of 0.2 
arcsec, and $\pm 0.02$ error at 0.4 arcsec. 
Uncertainties in image centroiding thus can not explain the difference between the 
$M'$ and shorter wavelength ellipticity measurements near 0.5 arcsec 
radius. It would be of interest to obtain additional images near $5\mu$m to 
investigate further the MIR ellipticities between 0.5 and 0.7 arcsec. 

	The isohotal properties of the $K'$ and $L'$ images, 
with the $K'$ image smoothed to the angular resolution of the $L'$ image, 
are compared in Figure 17. The angular resolution of the data used to construct 
Figure 17 is thus 0.2 arcsec, as opposed to the 0.4 arcsec resolution of the data 
used for Figure 16. The $K'$ and $L'$ surface brightness profiles have 
similar slopes, and this is reflected in the flat $K'-L'$ profile, shown in 
the middle panel. The $K'$ and $L'$ ellipticity measurements are also 
similar at radii $> 0.2$ arcsec. The tendency for ellipticity to decrease 
at radii $< 0.6$ arcsec is likely due -- at least in part -- to seeing effects 
(e.g. Erwin \& Sparke 2003). The comparisons in Figures 16 and 17 indicate 
that the `stubborn normalcy' in stellar content near the center of M32 
noted by Lauer et al. (1998) is evident at angular resolutions 0.2 -- 0.4 arcsec
in the NIR and MIR. 

\section{DISCUSSION \& SUMMARY}

	The stellar content of the Local Group cE galaxy M32 has been investigated 
using ground and space-based images that span the $1 - 8\mu$m wavelength 
region. A goal of this study is to use M32 as a foil to 
consider the utility of MIR observations for investigating the stellar contents of 
more distant galaxies. M32 is a favorable target for checking models and developing 
techniques that might be applied to trace the evolution of more distant galaxies 
as it is relatively nearby, allowing stars near 
the MSTO to be resolved, and has been extensively studied. 

	It is demonstrated that accounting for the LPV nature of bright 
AGB stars in M32 is an essential element of successfully 
modelling the NIR and MIR photometric properties of these objects. In the current study 
this has been done by convolving model LFs with an amplitude distribution 
that is based on Galactic LPVs. While variability smooths out 
features in the LF, it does not completely obliterate information that can be 
used to probe the SFH. Indeed, the overall shape of the smoothed [5.8] LF can be 
used to investigate the contributions made by stars with intermediate and old ages. 
Comparisons between the NIR and MIR LFs and models constructed from the Bressan 
et al. (2012) sequences indicate that M32 is not a SSP, and must contain an 
old, metal-rich stellar component. Model $K$ and [5.8] LFs that 
assume the SFH found by Monachesi et al. (2012) match the observations in both filters. 
The potential limitations imposed by stellar variability notwithstanding, the results 
in this paper indicate that observations of more distant systems in the MIR, such as 
those that will be conducted with facilities such as the JWST, will prove useful for 
constraining the fraction of old and intermediate age stars in galaxies where other 
moderately bright age indicators, such as core helium burning stars, can not be resolved.

	It would be of interest to investigate the 
NIR and MIR LFs of Local Group galaxies that contain AGB stars 
with ages and/or metallicities that differ from those in M32. 
Models predict that the rate of mass loss on the 
upper AGB, and hence the size of circumstellar envelopes, depends on 
progenitor mass and metallicity (e.g. Bowen \& Willson 1991), although there 
is still not extensive observational evidence to support a metallicity-dependence 
(e.g. Lagadec 2010). The models predict that more massive, metal-rich 
AGB stars will have thicker, dustier circumstellar envelopes than 
less massive, more metal-poor objects, and the former are expected to be brighter 
in the MIR than the latter. Hence, a different fraction of the MIR 
light might be expected to originate from circumstellar sources 
in galaxies with mean metallicities that differ from that of M32.

	We close the paper by briefly considering the origins of M32, as the 
analysis of the NIR and MIR data provides information that is relevant to this issue. 
The NIR and MIR images suggest that M32 may be structurally complex. 
The isophotal analysis discussed in Section 6 reveals a flattened 
structure in the unresolved NIR and MIR light that can be 
traced to within 16 arcsec of the galaxy center. 
The coefficient of the fourth order cosine expansion of the MIR isophotes 
is also indicative of a disky morphology at $r > 48$ arcsec. 
These results are interesting given that Graham (2002) and Choi et al. (2002) 
find evidence for a diffuse disk-like distribution in the outer regions of M32. 

	If a disk is present then clues to its origins will be 
found in its stellar content. For example, a 
diffuse remnant of a once-larger progenitor disk might be expected to show 
a SFH that was truncated at the time of the events that disrupted 
the disk, and there might be evidence for elevated galaxy-wide 
star-forming activity. On the other hand, a diffuse disk that 
formed during cosmologically recent epochs from the accretion 
of high angular momentum gas that was able to cool and form 
stars might show a star-forming history that is distinct from the main body 
of the galaxy, and that continued to relatively recent epochs.

	The changes in structural characteristics found here are not 
accompanied by an obvious radial change in stellar content. There 
are no large-scale gradients in absorption line 
strengths in M32, although fortuitous gradients in age and metallicity that conspire 
to negate absorption line gradients can not be 
ruled out (Davidge 1991; Worthey 2004; Rose et al. 2005). 
While in Section 4 it is shown that there is a radial gradient 
in the mean $J-K$ color of the M32 AGB sequence, this trend 
is consistent with progressively larger fractional contamination from stars in 
the outer disk of M31 towards larger distances from 
the center of M32, rather than a gradient in -- say -- 
mean metallicity within M32. More rigorous constraints 
on stellar content variations are placed by the specific frequency of bright 
AGB stars, which does not vary with radius out to very large radii when 
normalized according to either visible or NIR surface brightnesses (Davidge \& 
Jensen 2007). The peak $3.5\mu$m brightness of AGB stars in M32 also does not change 
over offsets from 2 arcsec to 176 arcsec from the galaxy center (Section 4), 
underscoring previous studies at shorter wavelengths that have found that the 
brightest AGB stars are distributed throughout the galaxy. If there is a diffuse 
disk around M32 then it has a stellar content that is not greatly different 
from that of the underlying body of the galaxy.

	Could M32 be a fossil remnant of a major merger? 
Hammer et al. (2010) investigate the possibility of a major merger between M31 
and another galaxy a few Gyr in the past. In addition to 
producing multiple tidal features around M31, such an event can 
also explain episodes of elevated star formation that are 
evident in the age distribution of M31 globular clusters (e.g. Puzia et al. 2005), 
and the diverse nature of the M31 globular cluster system (e.g. Beasley et al. 2005).
It has been demonstrated in this paper that the NIR and MIR photometric 
properties of AGB stars in M32 are consistent with significant extended periods of 
star-forming activity in M32 during intermediate epochs, and this 
is in qualitative agreement with the interaction modelled by Hammer et al. (2010). 
It would be of interest to determine if there are complex debris trails in 
the vicinity of M31 of the type predicted by the Hammer et al. (2010) simulations 
that are populated by AGB stars with NIR and MIR properties like those studied here.

\acknowledgements{Thanks are extended to the anonymous referee for a timely and 
comprehensive report that greatly improved the presentation of the results.}

\parindent=0.0cm

\clearpage

\begin{table*}
\begin{center}
\begin{tabular}{lccl}
\tableline\tableline
Filter & R$_e^1$ & $\mu_e^2$ & Reference \\
 & (arcsec) & (mag arcsec$^{-2}$) & \\
\tableline
B$+$I & 29 & 19.43 (B), 17.53 (I) & Choi et al. (2002) \\
R & 32 & 18.79 & Kent (1987) \\
J & $36.6 \pm 0.6$ & $16.92 \pm 0.01$ & This paper \\
Ks & $34.1 \pm 0.6$ & $15.86 \pm 0.01$ & This paper \\
$[3.6]$ & $33.6 \pm 0.7$ & $20.34 \pm 0.02$ & This paper \\
$[4.5]$ & $33.1 \pm 0.8$ & $20.41 \pm 0.02$ & This paper \\
\tableline
\end{tabular}
\tablenotetext{1}{Effective radius.}
\tablenotetext{2}{Effective surface brightness.}
\caption{Light Profile Measurements}
\end{center}
\end{table*}

\clearpage

\begin{table*}
\begin{center}
\begin{tabular}{lrr}
\tableline\tableline
Filter & $<B4>$ & $<B4>$ \\
 & (r $< 16$ arcsec) & (r $> 48$ arcsec) \\
$J$ & --0.0013 & 0.0024 \\
 & $\pm 0.0024$ & $\pm 0.0011$ \\
 & & \\
$Ks$ & --0.0045 & -0.0010 \\
 & $\pm 0.0026$ & $\pm 0.0013$ \\
 & & \\
$[3.6]$ & --0.0007 & 0.0024 \\
 & $\pm 0.0015$ & $\pm 0.0006$ \\
 & & \\
$[4.5]$ & --0.0006 & 0.0033 \\
 & $\pm 0.0022$ & $\pm 0.0006$ \\
 & & \\
\tableline
\tableline
\end{tabular}
\caption{Mean B4 Coefficients}
\end{center}
\end{table*}

\clearpage

\begin{figure}
\figurenum{1}
\epsscale{1.00}
\plotone{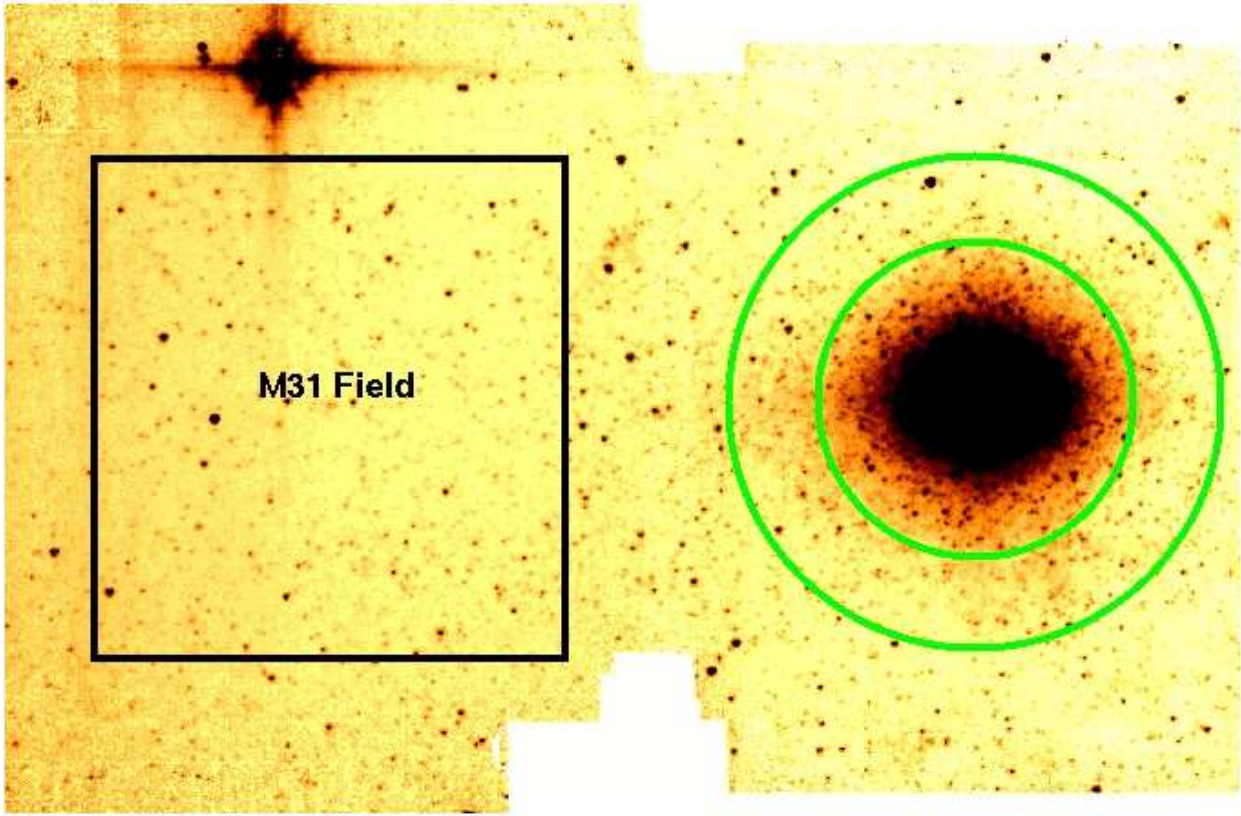}
\caption{The [5.8] PBCD image of M32. North is to the top and east 
is to the left. The Inner, Middle, and Outer annuli that are used to 
investigate the stellar content of M32 fall between the two green circles.
The region that is used to monitor the stellar content in the M31 outer 
disk is also indicated.} 
\end{figure}

\clearpage

\begin{figure}
\figurenum{2}
\epsscale{1.00}
\plotone{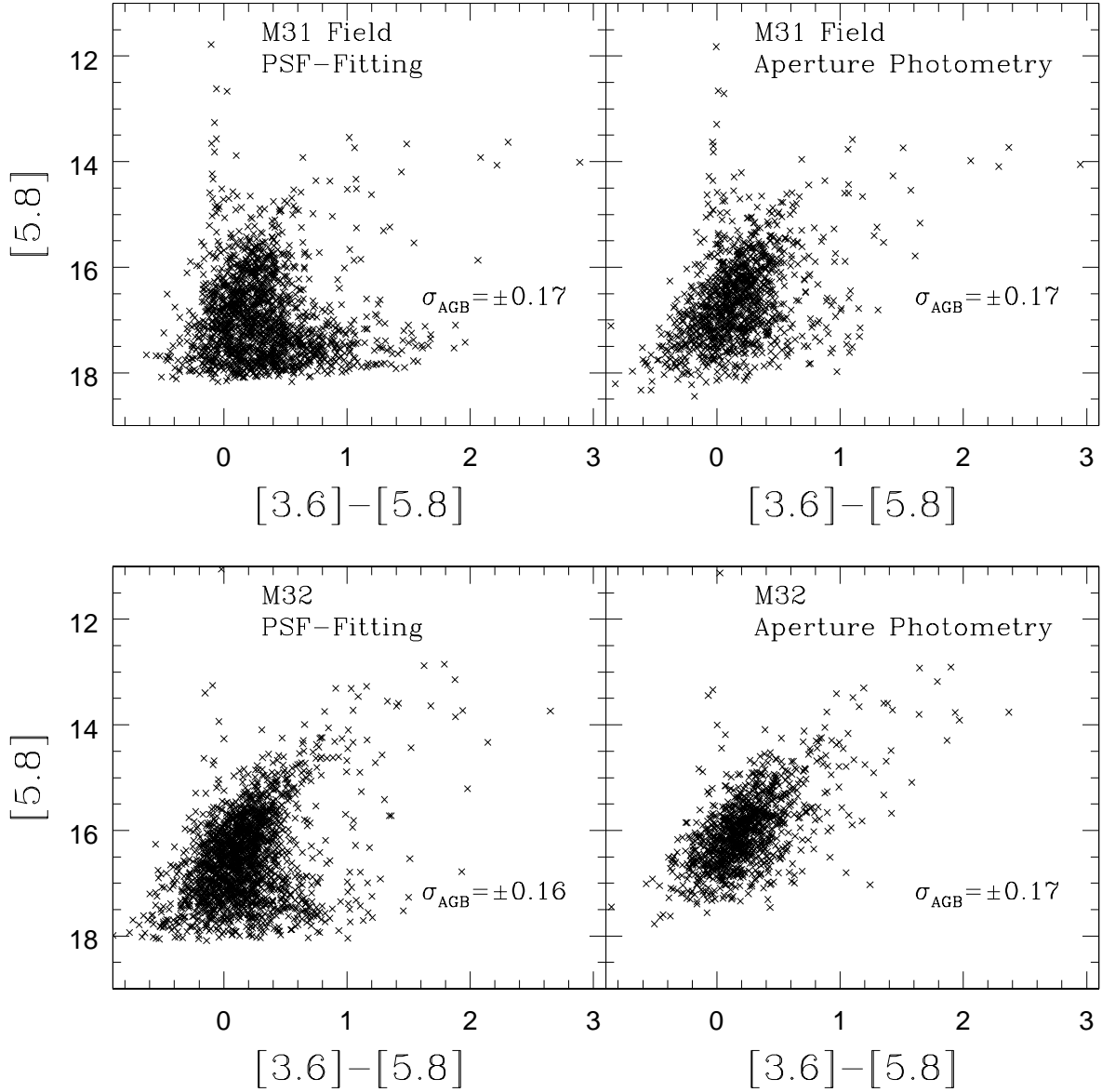}
\caption{$([5.8], [3.6]-[5.8])$ CMDs of the M31 outer disk (top row)
and M32 (bottom row), obtained with PSF-fitting and aperture 
photometry techniques. The standard deviation in [3.6] -- 
[5.8] about the mean [3.6]--[5.8] for stars with [5.8] between 15.75 and 16.25, 
$\sigma_{AGB}$, is shown for each CMD. The M32 AGB sequence that is obtained from 
PSF-fitting is tighter for sources with [5.8] $< 15$ than that obtained from 
aperture measurements. The CMDs constructed from PSF-fitting 
go deeper, have dispersions along the color axis that 
are comparable to or smaller than those obtained from aperture photometry, 
and produce better-defined sequences near the bright end.}
\end{figure}

\clearpage

\begin{figure}
\figurenum{3}
\epsscale{0.8}
\plotone{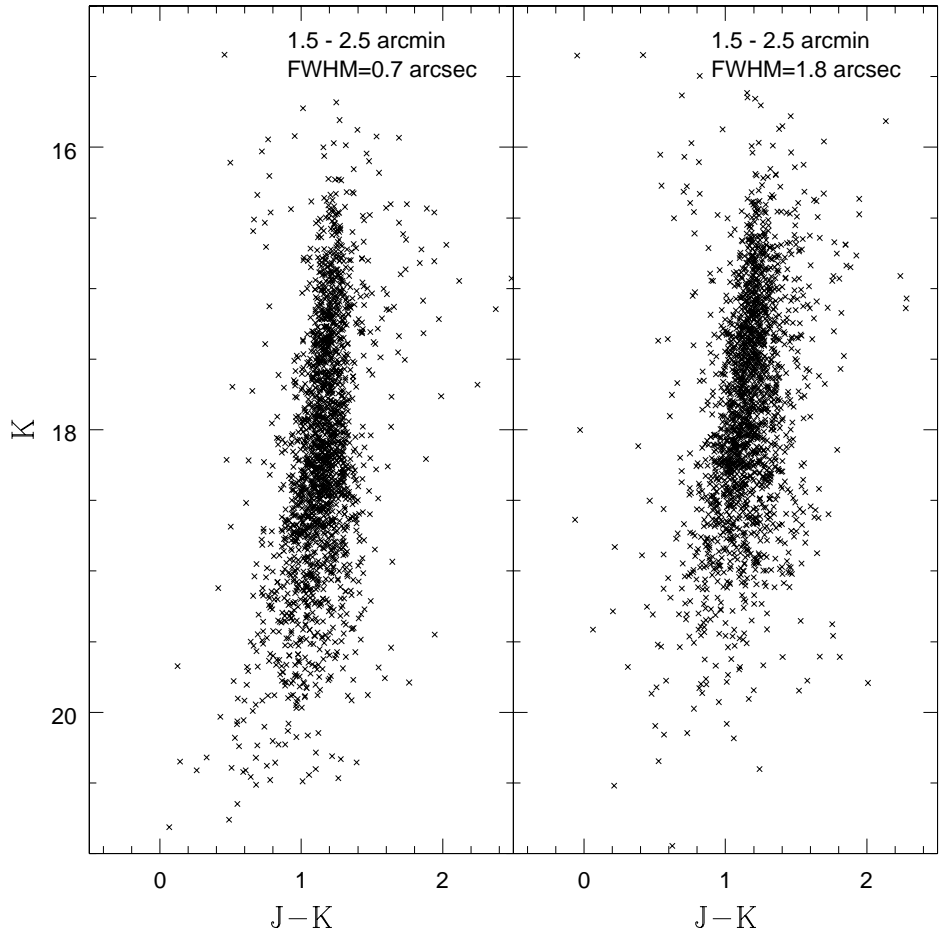}
\caption{$(K, J-K)$ CMDs of sources that are between 1.5 and 2.5 arcmin from the center 
of M32. The photometry used in the left hand CMD was obtained from the unsmoothed 
WIRCam observations, which have 0.7 arcsec FWHM angular resolution. The right 
hand CMD uses photometry obtained from the same images, but 
after they had been Gaussian-smoothed to match the 1.8 arcsec 
FWHM resolution of the SPITZER [3.6] images. As expected, 
smoothing elevates the faint limit of the CMD. However, the overall properties 
of the bright end of the AGB sequence are not drastically altered. Given that the 
increased contrast between AGB stars and the main body of stars in the MIR makes 
crowding even less of a concern in [3.6] (and the other IRAC passbands) than in 
$J$ and $Ks$, then these results suggest that crowding should 
not be an issue for the SPITZER data in the outer regions of M32. This is consistent 
with the the radial properties of stars in the MIR CMDs (Figure 6).}
\end{figure}

\clearpage

\begin{figure}
\figurenum{4}
\epsscale{1.00}
\plotone{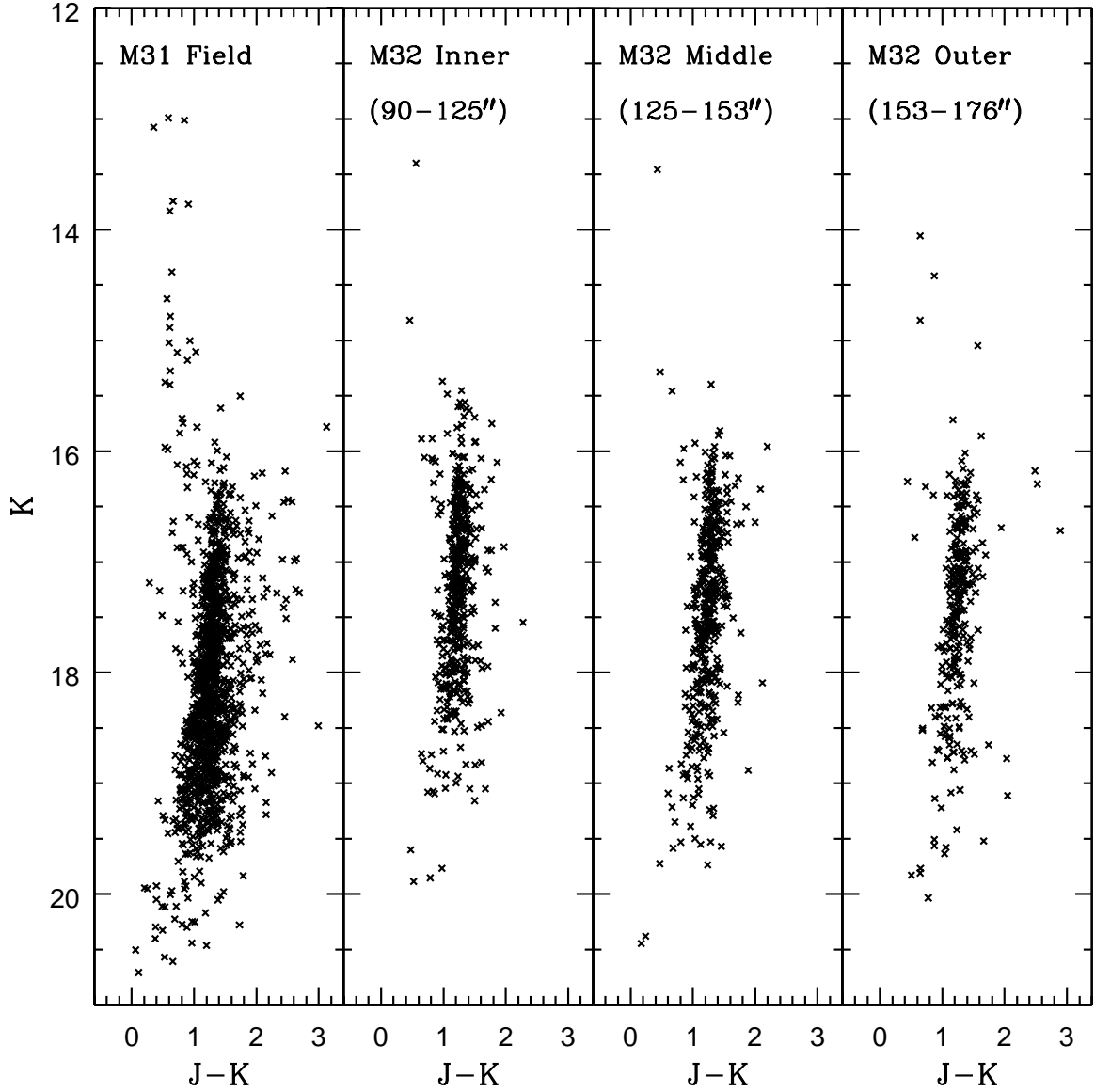}
\caption{$(K, J-K)$ CMDs of stars in M32 and the M31 outer disk. The M32 CMDs sample 
stars in three annuli centered on the galaxy and covering equal areas on the sky. 
The M31 CMD extends to fainter magnitudes than the M32 CMDs due to the lower 
density of sources in that field. The sequence with $J-K \sim 0.6$ in the M31 Field 
CMD is populated by K supergiants (Figure 5).}
\end{figure}

\clearpage

\begin{figure}
\figurenum{5}
\epsscale{1.00}
\plotone{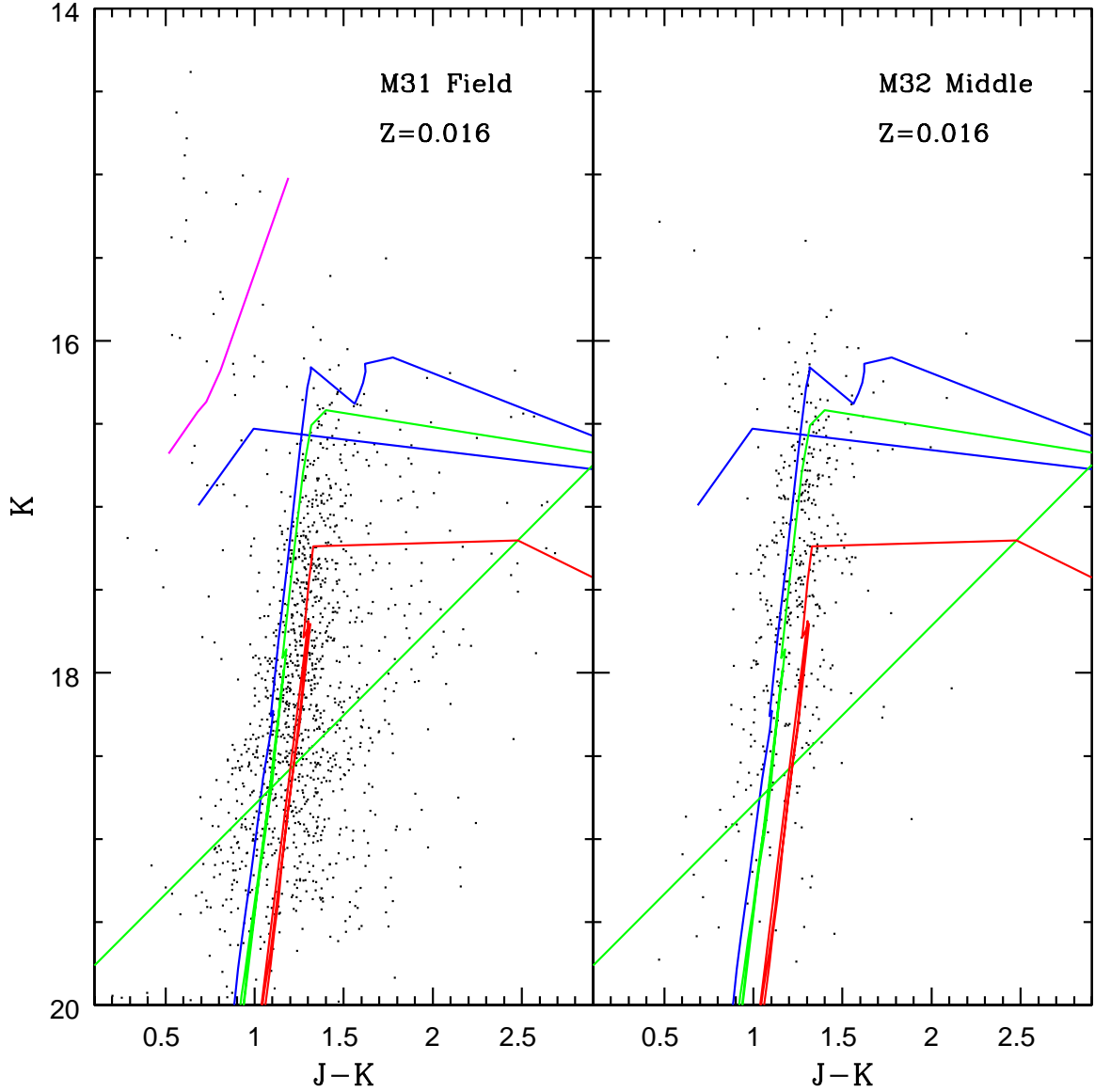}
\caption{Comparisons with isochrones from Marigo et al. (2008). Sequences with Z=0.016 
and ages 1 (blue), 3 (green), and 10 (red) Gyr are shown. The magenta 
sequence in the left hand panel is the locus of Galactic G and K RSGs from 
Koornneef (1983).}
\end{figure}

\clearpage

\begin{figure}
\figurenum{6}
\epsscale{1.00}
\plotone{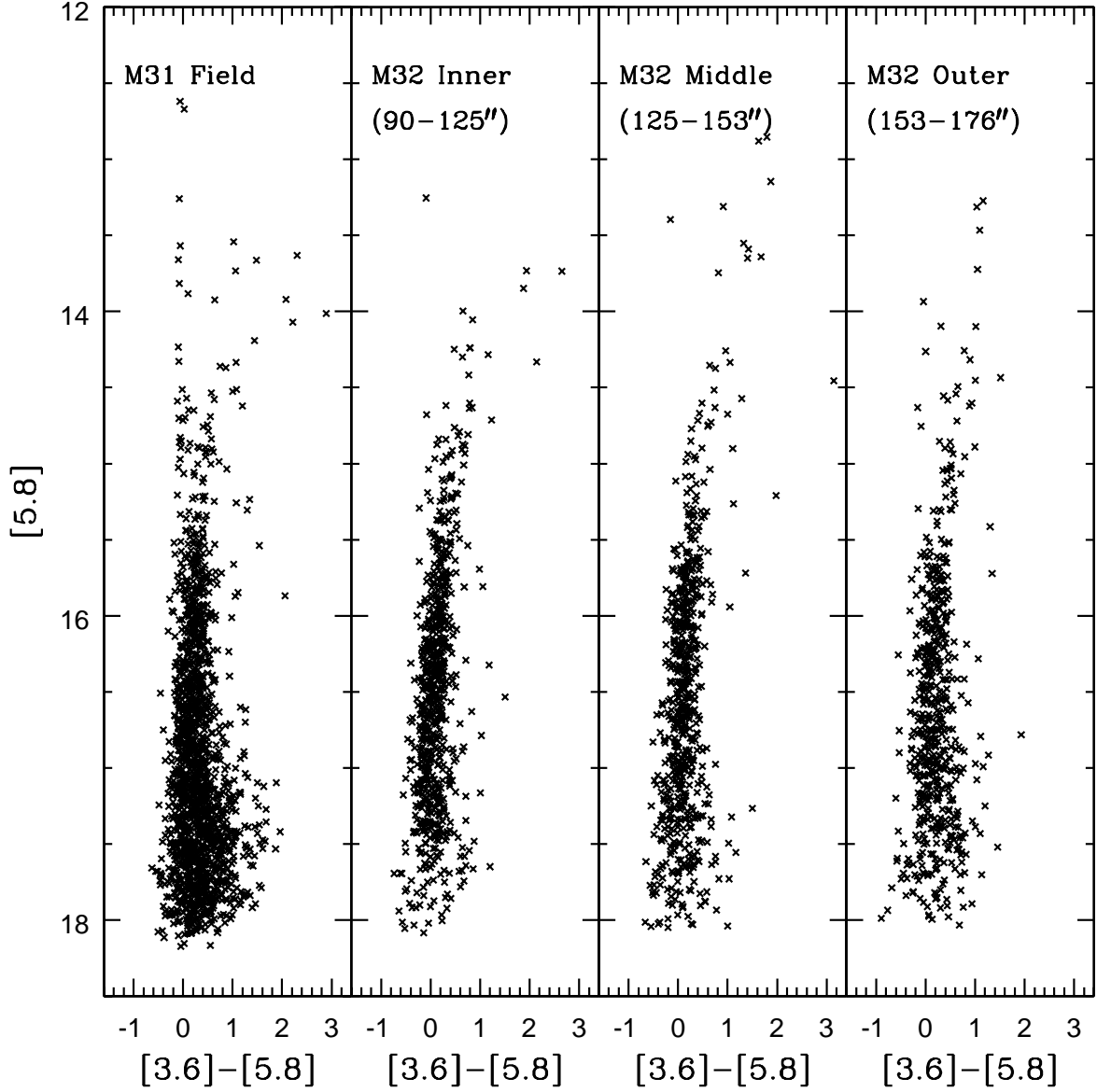}
\caption{$([5.8], [3.6]-[5.8])$ CMDs. The majority of objects in the CMDs are 
AGB stars. Objects with $[5.8] < 15.5$ have $[3.6]-[5.8] > 0$ 
suggesting significant thermal emission from circumstellar dust. The 
sequence with $[3.6]-[5.8] \sim 0$ and $[5.8] < 14.5$ 
in the left hand panel is populated by foreground Galactic dwarfs, members 
of M31 and M32, and background galaxies. This sequence is also present in the M32 CMDs, 
although it is not as well populated due to the smaller angular coverage of each 
annulus.}
\end{figure}

\clearpage

\begin{figure}
\figurenum{7}
\epsscale{1.00}
\plotone{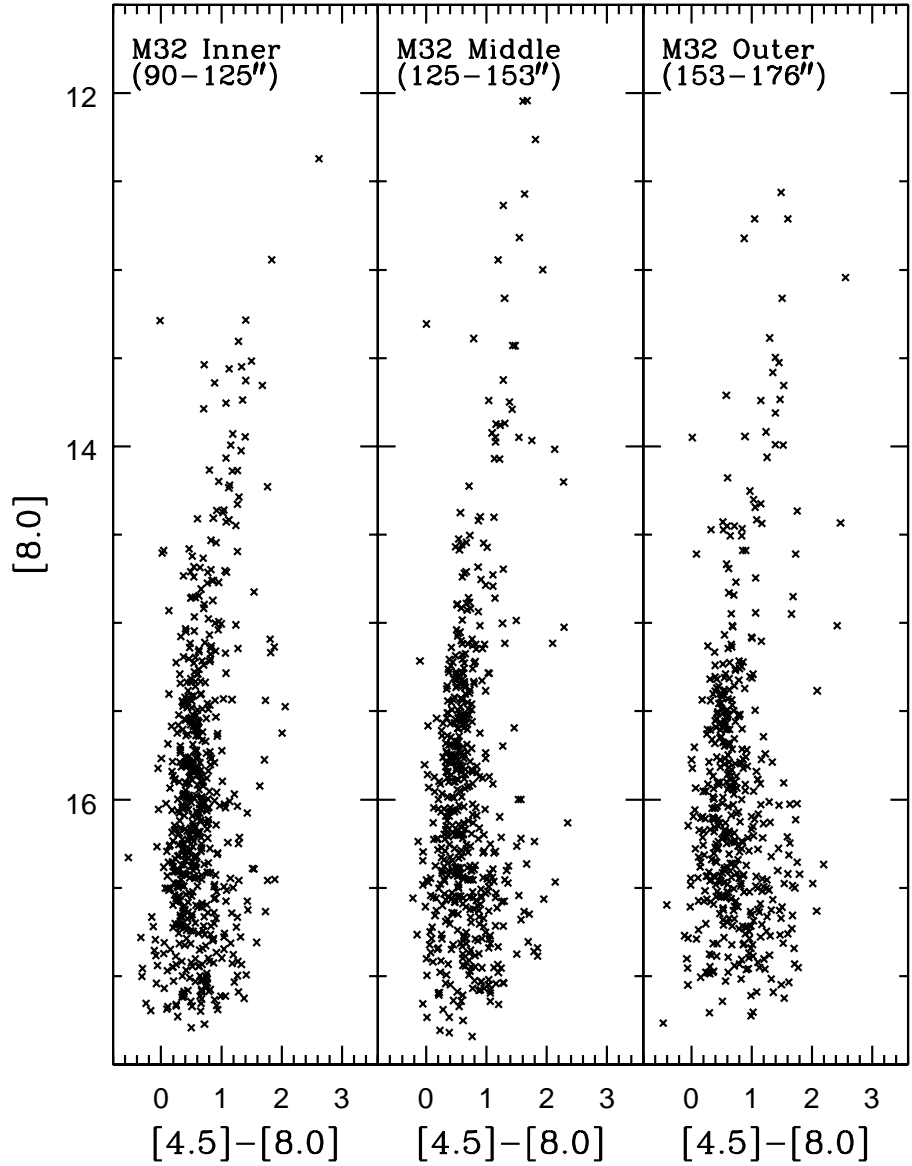}
\caption{$([8.0], [4.5]-[8.0])$ CMDs. The majority of objects are evolving on 
the AGB. The trend to larger $[4.5]-[8.0]$ colors towards brighter [8.0] 
magnitudes reflects an increasing contribution from circumstellar dust emission.}
\end{figure}

\clearpage

\begin{figure}
\figurenum{8}
\epsscale{1.00}
\plotone{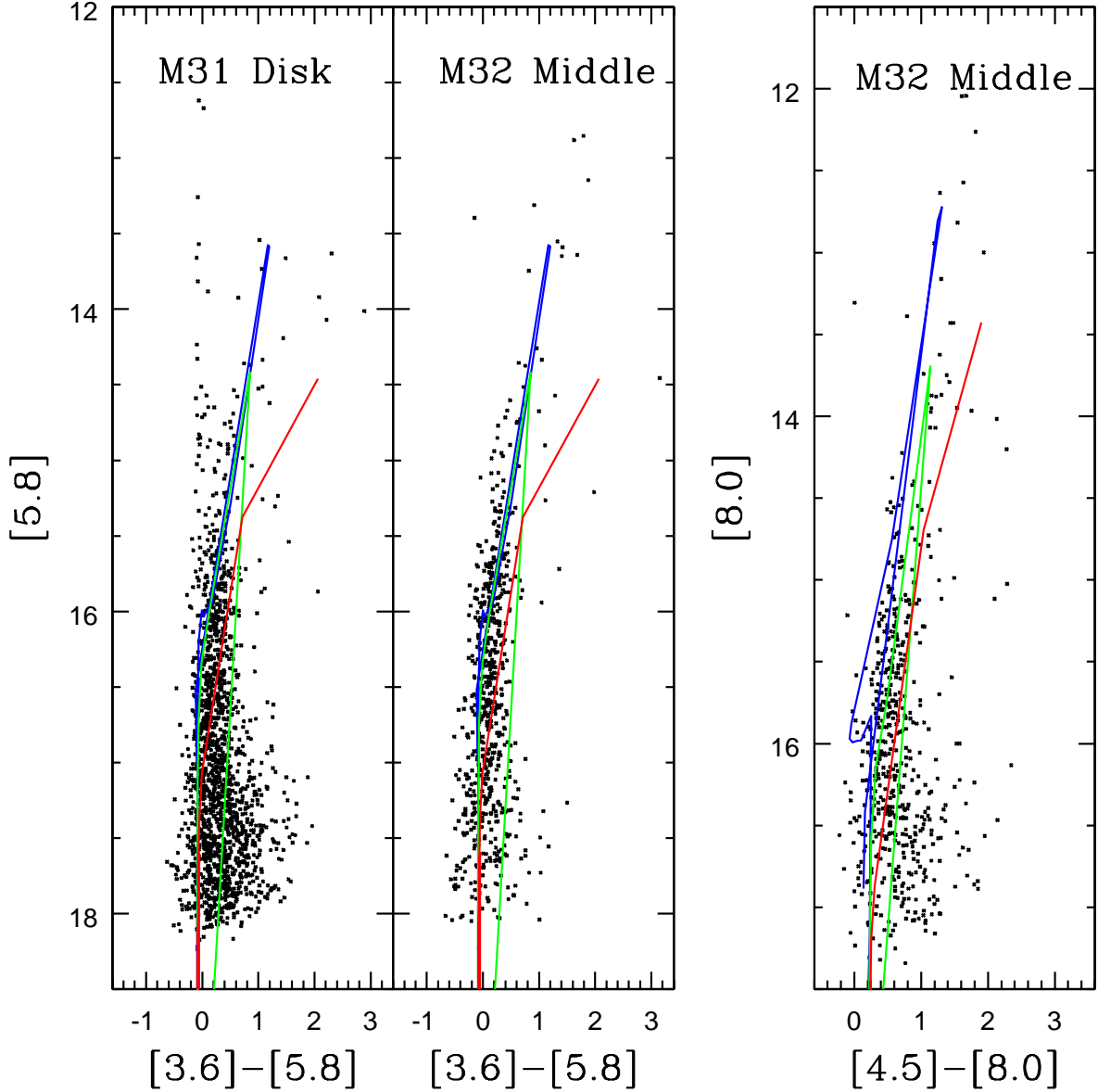}
\caption{Comparisons with Z=0.016 isochrones from Marigo et al. (2008). Dust in 
the circumstellar envelope is assumed to be 40\% ALOx $+ 60\%$ silicate. 
Sequences with ages 1 Gyr (blue), 3 Gyr (green), and 10 Gyr (red) are shown. 
The LPV nature of many of the brightest AGB stars 
complicates efforts to match their properties. 
Still, the $([5.8], [3.6]-[5.8])$ CMDs suggest that M32 and the M31 outer 
disk contain objects that span a range of ages. The 1 Gyr and 3 Gyr models track 
the brightest stars in the $([5.8],[3.6]-[5.8])$ CMDs of the M31 
outer disk and the M32 Middle annulus, although the models underestimate 
the peak brightness of stars in M32. Stars at intermediate magnitudes in the 
$([8.0], [4.5]-[8.0])$ CMD have $[4.5]-[8.0]$ colors that are consistent with 
an age $\sim 3$ Gyr.}
\end{figure}

\clearpage

\begin{figure}
\figurenum{9}
\epsscale{1.00}
\plotone{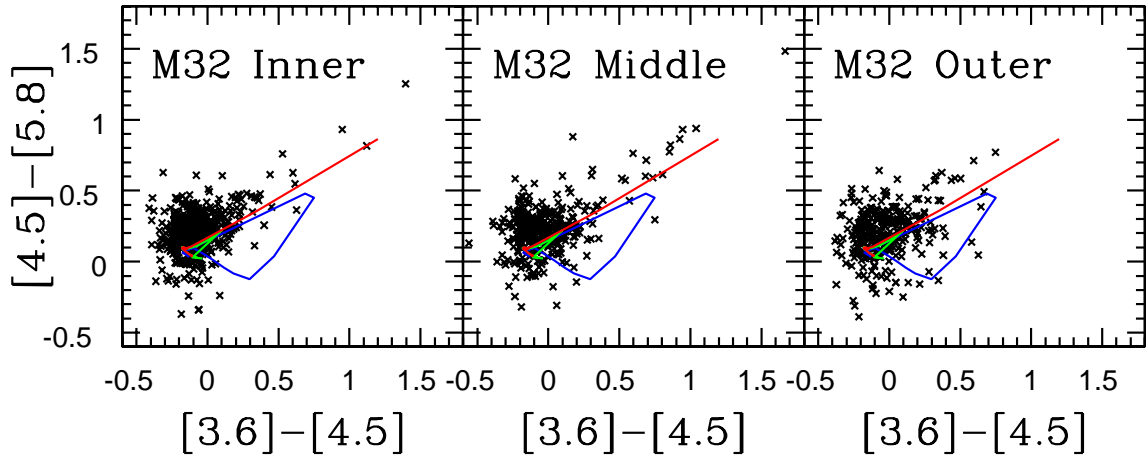}
\caption{$([3.6]-[4.5],[4.5]-[5.8])$ TCDs of objects with $[4.5] < 17$. 
Post-RGB sequences from Marigo et al. (2008) with Z=0.016 and ages 1 Gyr (blue), 3 Gyr 
(green), and 10 Gyr (red) are shown. The isochrones mimic the general trends 
defined by the data, although the red sequences predicted by the models
fall $\sim 0.1 - 0.2$ magnitudes below the observations.}
\end{figure}

\clearpage

\begin{figure}
\figurenum{10}
\epsscale{0.75}
\plotone{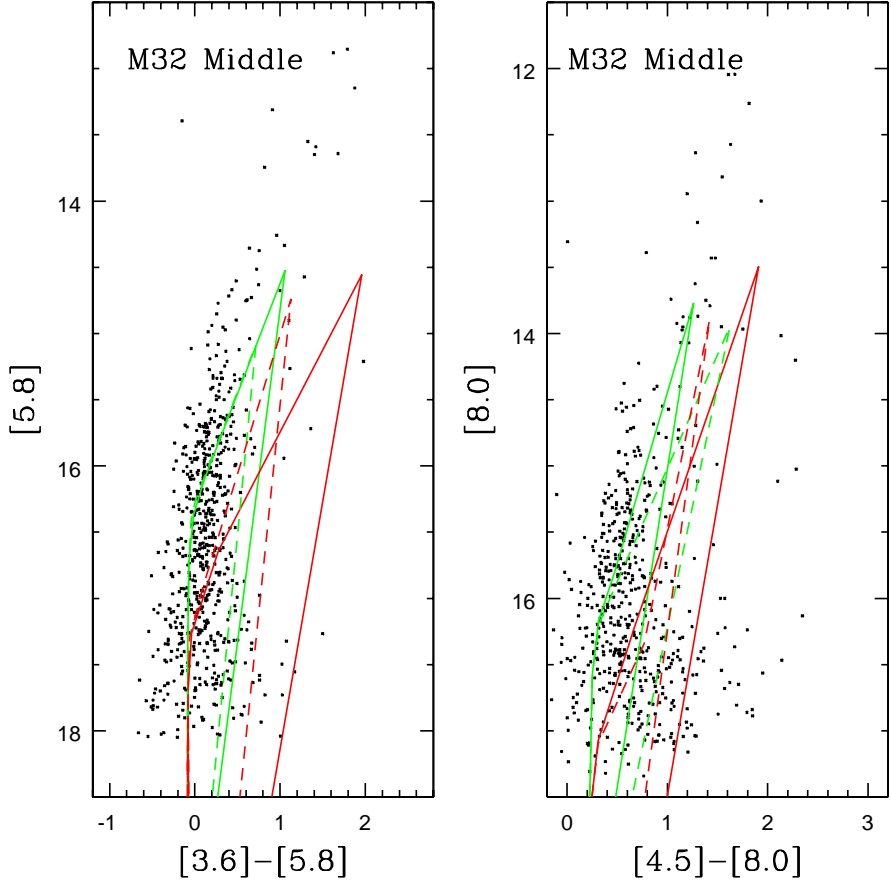}
\caption{Isochrones with different circumstellar chemical compositions. 
Sequences with Z=0.016 from Marigo et al. (2008) and ages 3 Gyr (green) and 10 
Gyr (red) are shown for two dust compositions: 100\% silicates (solid line), and 
100\% ALOx (dashed lines). The mean color of the 3 Gyr isochrone on the 
$([5.8],[3.6]-[5.8])$ CMD is similar for both compositions. While the 10 Gyr 
model on the $([5.8],[3.6]-[5.8])$ CMDs is sensitive to the composition of 
circumstellar dust, the basic conclusion that M32 contains a large 
population of intermediate age stars is independent of the assumed dust chemistry. 
The peak brightnesses of the model sequences on the 
$([8.0],[4.5]-[8.0])$ CMD are also not susceptible to dust composition. 
The 3 and 10 Gyr models with a 100\% ALOx composition have similar colors and fall 
redward of the ridgeline of the M32 sequence on that CMD, indicating that 
at least some silicates must be present in circumstellar dust in M31 and M32.}
\end{figure}

\clearpage

\begin{figure}
\figurenum{11}
\epsscale{0.75}
\plotone{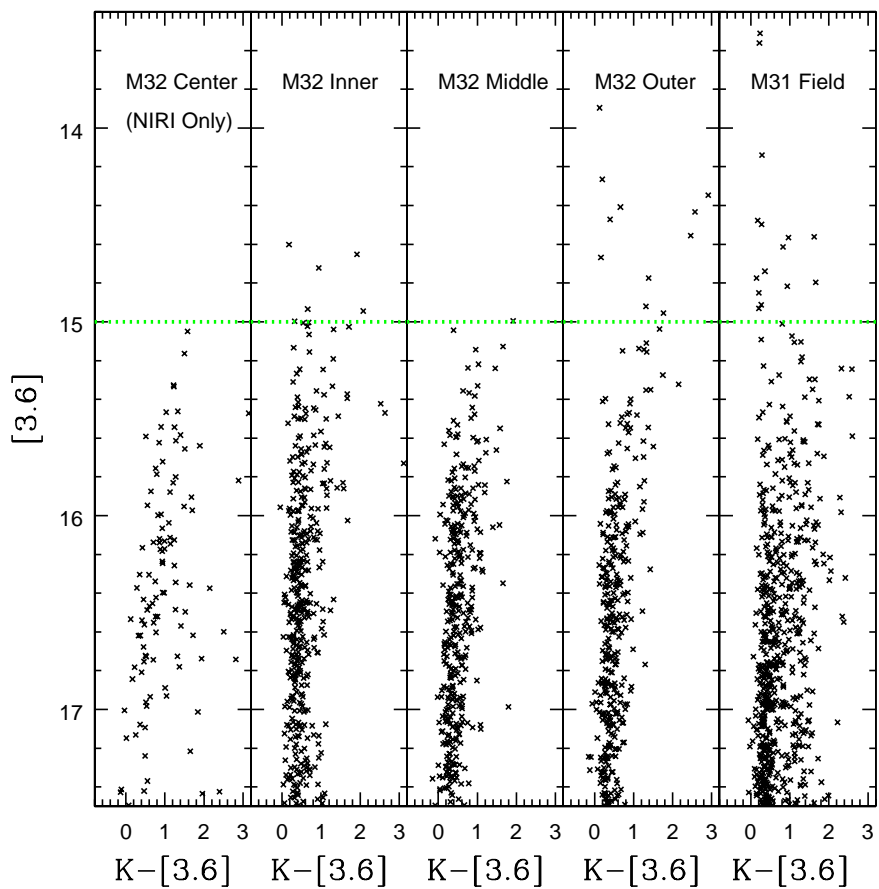}
\caption{$([3.6],K-[3.6])$ CMDs. The CMD in the left hand panel 
uses photometry obtained from the NIRI images, with the $L'$ NIRI 
measurements transformed into the IRAC [3.6] system. 
The CMDs in other panels use photometry obtained from 
the IRAC [3.6] and WIRCam $K$ images. The faint limit of these data 
corresponds to the approximate magnitude of the RGB-tip in M32. The dotted 
green line marks [3.6] = 15, which is the approximate peak brightness in the Center 
field.} 
\end{figure}

\clearpage

\begin{figure}
\figurenum{12}
\epsscale{1.00}
\plotone{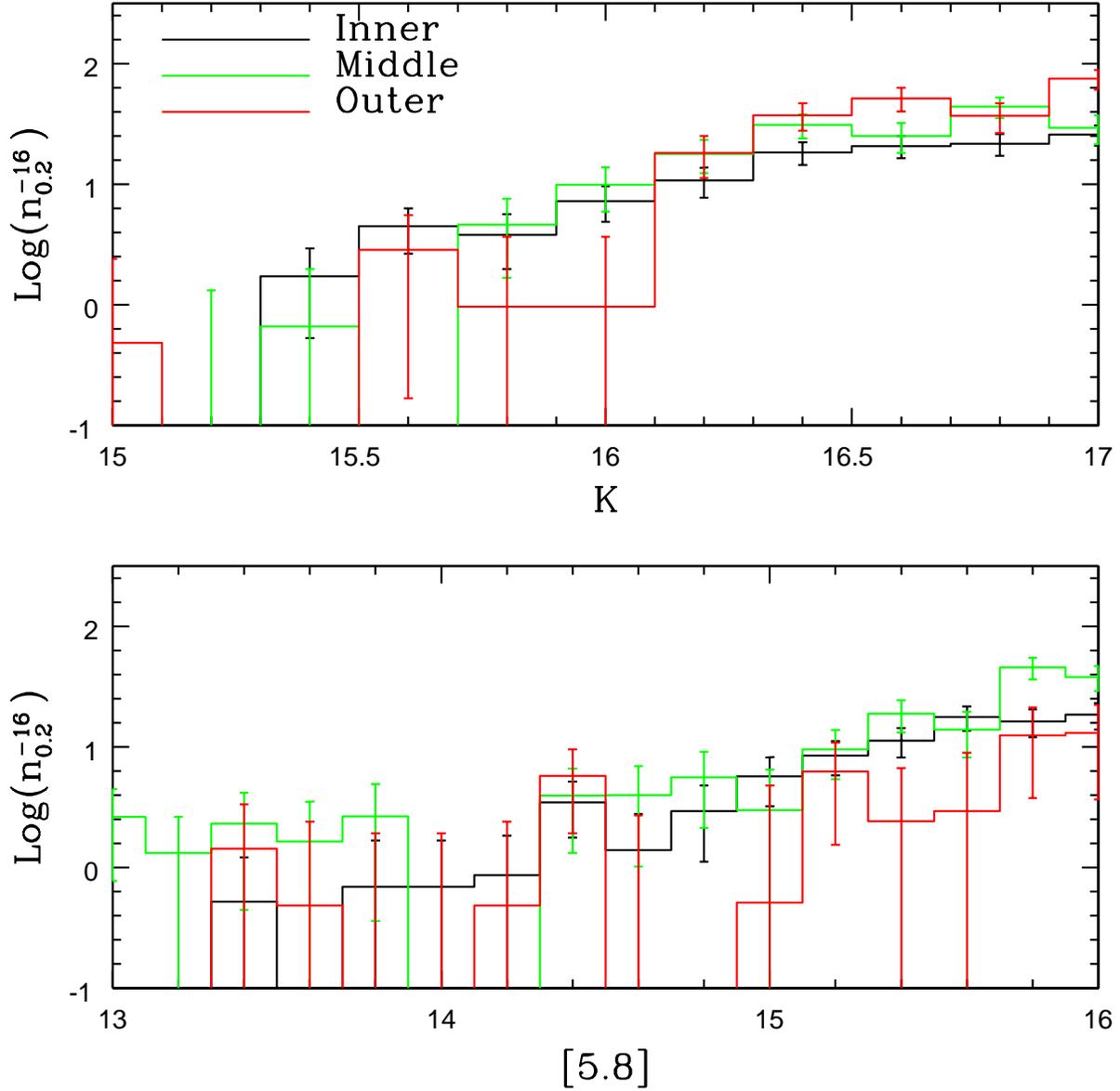}
\caption{$K$ and [5.8] LFs. n$^{-16}_{0.2}$ is the specific 
frequency of objects per 0.2 magnitude interval 
in a system with M$_K = -16$. The LF of the M31 outer disk, 
scaled to match the area sampled in each annulus, was subtracted from 
the M32 LFs to remove objects in the foreground, stars in the disk of M31, and 
background galaxies. The error bars are $1\sigma$ uncertainties based on 
counting statistics. The specific frequencies of stars in all three annuli agree 
to within the uncertainties, although there is a tendency for the [5.8] LF 
of the outer annulus to fall systematically below those of the inner and middle annuli.}
\end{figure}

\clearpage

\begin{figure}
\figurenum{13}
\epsscale{0.65}
\plotone{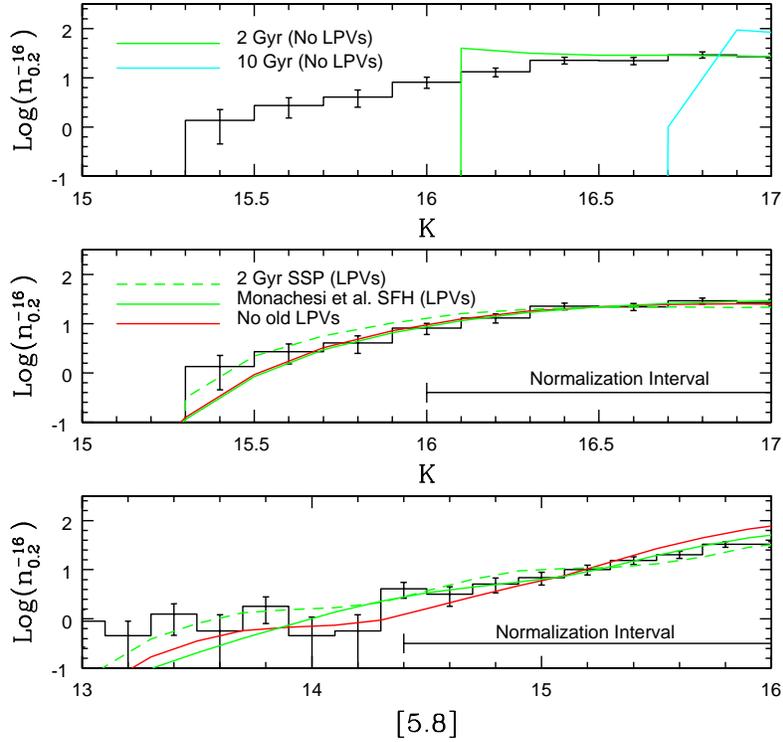}
\caption{Mean $K$ and [5.8] LFs of the inner and middle 
annuli are compared with models. n$^{-16}_{0.2}$ is the number of objects 
per 0.2 magnitude interval in a system with M$_K = -16$ after subtracting the 
M31 disk LF. (Top panel:) Model SSP LFs constructed from the Marigo et al. 
(2008) Z=0.016 isochrones that do not consider photometric variability. These models 
have been normalized to match the number counts in the last two bins of each LF. 
There is poor agreement with the observations. (Middle 
and lower panels:) Models convolved with the amplitude 
distribution function defined by Galactic bulge LPVs observed by Glass et al. 
(1995). When compared with the results in the top panel, 
it is clear that the addition of variability has a significant 
impact on the agreement between the 2 Gyr model LF and the observations. The dashed 
green lines in these panels are 2 Gyr SSP models, while the solid green 
lines are models that follow the SFH shown in Figure 13a of Monachesi et al. (2012). 
The models have been normalized to the observations in the magnitude intervals 
that are indicated. The composite model LF provides a better match to the $K$ 
and [5.8] observations than the 2 Gyr SSP model, although the composite models 
underestimate the numbers of bright AGB stars. The red line is a model that follows the 
Monachesi et al. (2012) SFH, with the exception that there are no stars older than 
7 Gyr. While such a model can explain the deficiency of bright AGB stars, 
this is done at the expense of the agreement at fainter [5.8] magnitudes.}
\end{figure}

\clearpage

\begin{figure}
\figurenum{14}
\epsscale{0.75}
\plotone{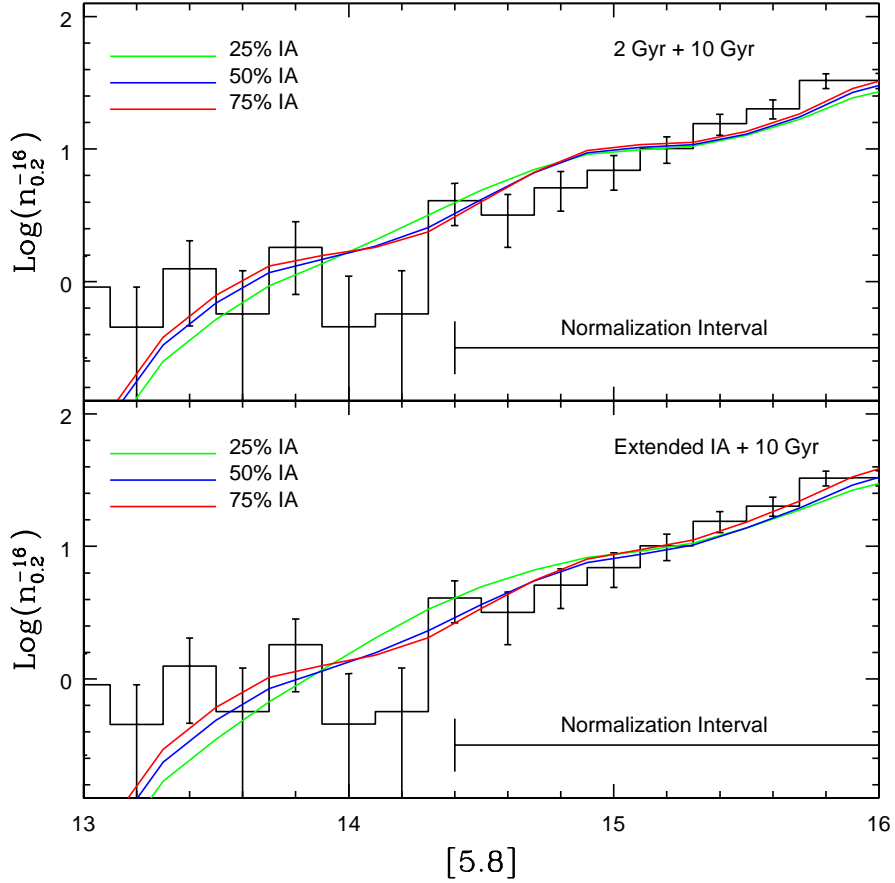}
\caption{Two-component model LFs. (Top panel:) Models in 
which a 2 Gyr SSP is combined with a 10 Gyr SSP, with three different 
stellar mass fractions for the intermediate age population, are compared 
with the [5.8] LF of the inner and middle annuli of M32. The models assume Z = 0.016. 
Note that these models do not match the slope of the observed LF between [5.8] 
= 14.5 and 16. (Bottom panel:) The same as the top panel, but with the 2 Gyr SSP 
replaced with a constant SFR event covering 2 -- 3 Gyr in the past. The models with 
an extended star-forming event provide a better match to the observed LF in the 
interval [5.8] = 14.5 to 16 than the discreet burst models. While an extended period of 
star formation during intermediate epochs is favored by the data, only weak 
constraints can be placed on the size of the intermediate age component.}
\end{figure}

\clearpage

\begin{figure}
\figurenum{15}
\epsscale{0.6}
\plotone{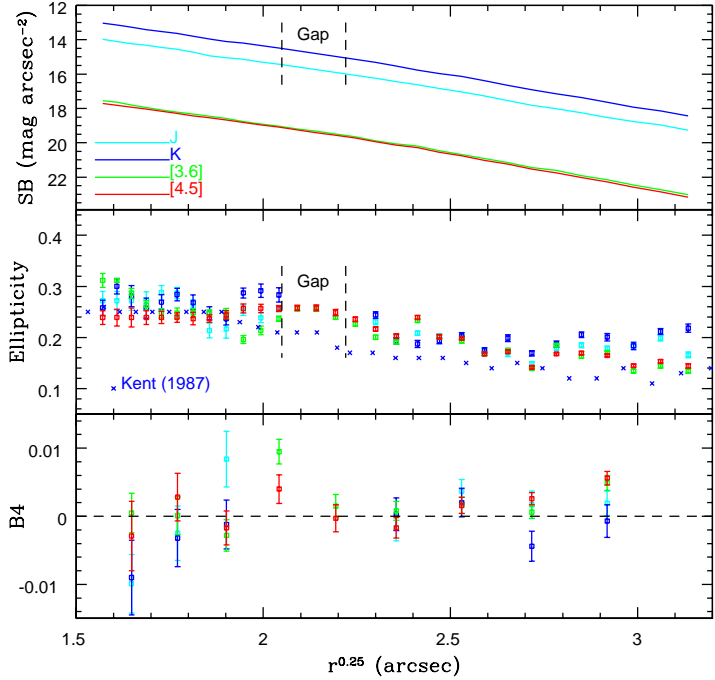}
\caption{Isophotal properties. (Top panel:) Light profiles. 
The gap in the radial coverage of the WIRCam observations due to 
a group of bad pixels is indicated. The agreement between the 
profile slopes indicates that the light distributions between 1 and $4.5\mu$m 
have similar R$_e$'s. The random uncertainties in the light profiles computed 
by $ellipse$ are $< 1\%$. (Middle panel:) Ellipticity measurements. The error 
bars show $1\sigma$ uncertainties that are computed by {\it ellipse} 
based on the scatter about the fitted isophote. The blue crosses are ellipticities 
measured by Kent (1987) from $R-$band images. While there is broad filter-to-filter 
agreement within the estimated uncertainties for radii $<16$ arcsec, between 16 and 
45 arcsec the ellipticities measured by Kent (1989) fall 
systematically below those at longer wavelengths, suggesting that the light in 
the visible/red is less flattened than in the NIR and MIR. Considering the stellar 
types that contribute at various wavelengths (e.g. Maraston 2005), this suggests that 
the spatial distributions of MSTO/SGB/RGB objects, which dominate the integrated light 
at visible wavelengths, differ from those of the evolved stars that contribute 
significantly to the light at longer wavelengths. (Lower panel:) Coefficients 
of fourth-order cosine terms in a fourier expansion of the isophotes. The 
measurements are three isophote averages, while the error bars are 
computed from the $1\sigma$ uncertainties calculated by $ellipse$. 
At $r < 16$ arcsec there is a tendency for B4 $< 0$, indicating a 
boxy morphology. However, for $r > 48$ arcsec B4 $> 0$ in $J$, [3.6], 
and [4.5] at more than the $2\sigma$ level (see Table 2), indicating a disky 
morphology. Note that the B4 measurements made in $Ks$ differ significantly from those 
in the other filters.}
\end{figure}

\clearpage

\begin{figure}
\figurenum{16}
\epsscale{0.75}
\plotone{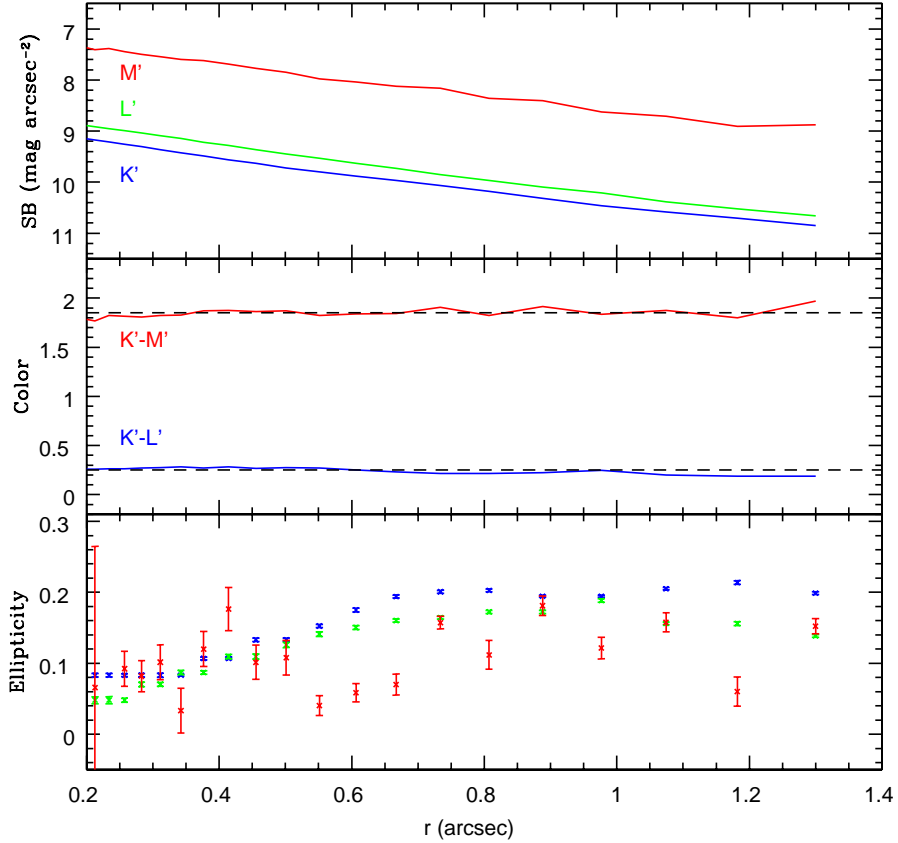}
\caption{Isophotal properties measured from NIRI images. 
The $K'$ and $L'$ images were convolved with Gaussians to match 
the 0.4 arcsec FWHM angular resolution of the $M'$ data. 
The isophotal properties in the central 0.2 arcsec are not shown given the 
diverse nature of the PSFs in these filters and the impact that even subtle differences 
in the PSF can have on central light profiles. (Top panel:) Light profiles. The 
uncertainties in the surface brightness measurements made by $ellipse$ are $\leq 1\%$, 
and so are not shown. (Middle panel:) $K'-L'$ and $K'-M'$ colors. The dashed lines 
mark the mean $K'-M'$ and $K'-L'$ colors. As with the surface brightness 
profile, the random uncertainties in the colors are modest, as might be expected 
given the absence of jitter in the color profiles. (Lower panel:) Ellipticities. 
The isophotes in the $K'$ and $L'$ images show similar amounts of 
flattening. Note that the ellipticities measured from the $M'$ image between 
0.5 and 0.7 arcsec differ from those found at shorter wavelengths.}
\end{figure}

\clearpage

\begin{figure}
\figurenum{17}
\epsscale{1.00}
\plotone{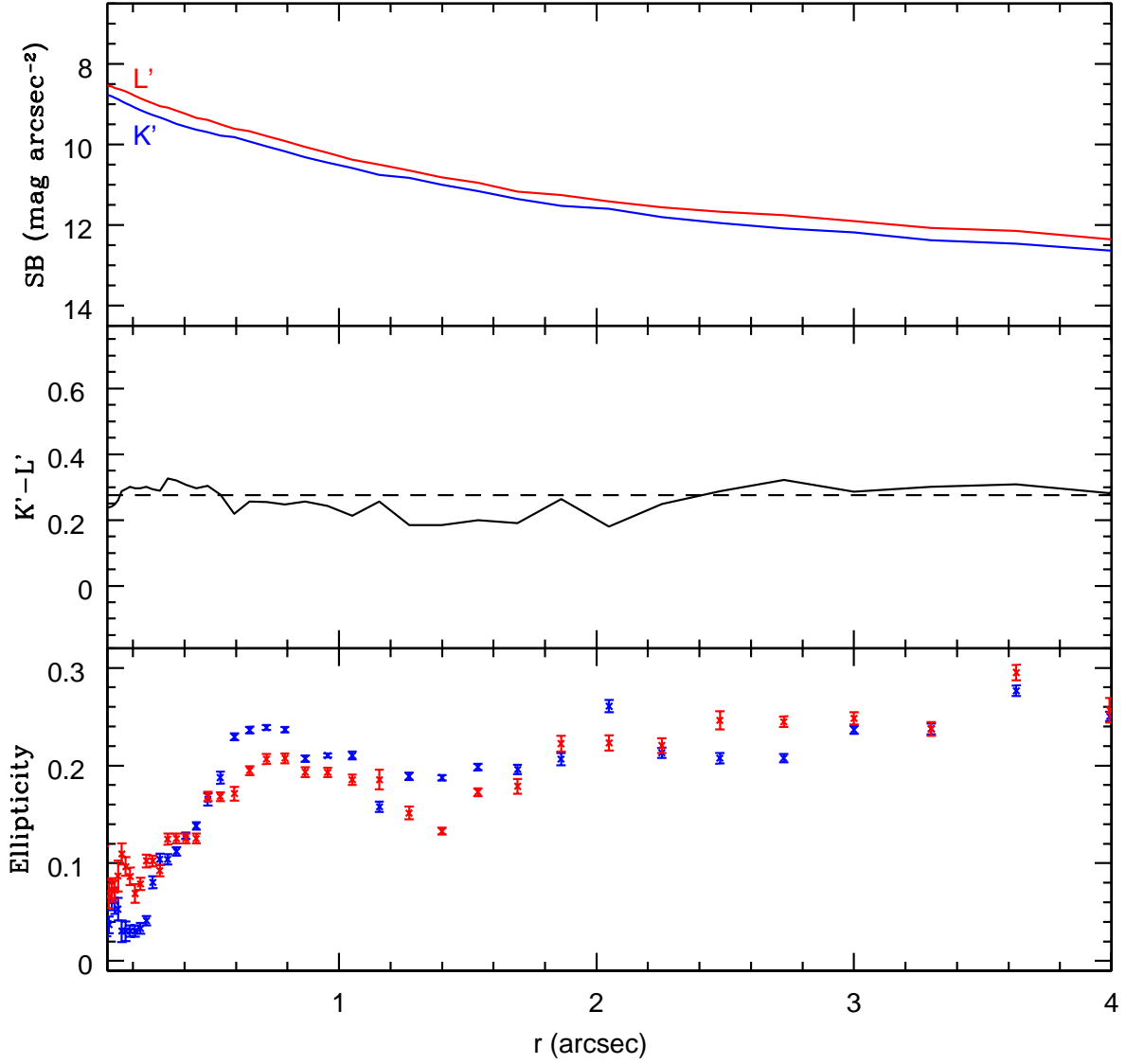}
\caption{Isophotal properties in $K'$ and $L'$. The $K'$ image has been 
Gaussian smoothed to match the angular resolution of the $L'$ image. (Top panel:)
Light profiles. The random uncertainties in the surface brightness measurements 
computed by $ellipse$ are $< 1\%$, and so are not shown here. 
(Middle panel:) $K'-L'$ colors. The dashed line indicates the mean $K'-L'$ color. 
The random uncertainties in $K'-L'$ are too small to show on this scale. 
(Bottom panel:) Ellipticities. The light profiles and ellipticities in $K'$ 
and $L'$ are similar in the central 4 arcsec of the galaxy.}
\end{figure}

\end{document}